\documentclass[12pt]{article}
\usepackage{graphicx}
\usepackage[margin=1.25in]{geometry}
\usepackage[usenames,dvipsnames]{color}
\usepackage{url}
\usepackage{caption,subcaption, cite}
\usepackage[colorlinks = true,
            linkcolor = blue,
            urlcolor  = blue,
            citecolor = blue,
            anchorcolor = blue]{hyperref}

\newcommand\pubnumber{DESY 19-146, IFIC/19-35\\ 
 KEK Preprint 2019-22 \\
 SLAC-PUB-17467}
\newcommand\pubdate{August, 2019}

\def\KEK{High Energy Accelerator Research Organization (KEK), Tsukuba,
  Ibaraki, JAPAN  }
\def\Tokyo{ICEPP, University of Tokyo, Hongo, Bunkyo-ku, Tokyo,
  113-0033, JAPAN}
\def\SNU{Dept. of Physics and Astronomy, Seoul National
  Univ.,  Seoul 08826, KOREA}
\def\DESY{DESY, Notkestrasse 85, 22607 Hamburg, GERMANY}
\def\Berlin{Institut f\"ur Physik, Humboldt-Universit\"at zu Berlin, 12489 Berlin, GERMANY}
\def\SLAC{SLAC,
    Stanford University, Menlo Park, CA 94025, USA}

\def\Peking{Department of Physics, Peking University, Beijing 100871, CHINA}
\def\Osaka{Department of Physics, Osaka University, Machikaneyama, Toyonaka, Osaka 560-0043, JAPAN}
\def\IPMU{Kavli Institute for the Physics and Mathematics of the Universe,
University of Tokyo, Kashiwa 277-8583, JAPAN}
\def\Cornell{Laboratory for Elementary Particle Physics, Cornell
  University, Ithaca, NY 14853, USA}
\def\Orsay{LAL, Centre Scientifique d'Orsay, Universit\'e Paris-Sud, F-91898 Orsay CEDEX,
FRANCE}
\def\Munich{Max-Planck-Institut f\"ur Physik, F\"ohringer Ring 6,
  80805 Munich, GERMANY}
\def\Michigan{Michigan Center for Theoretical Physics, University of Michigan, Ann Arbor,
MI 48109, USA}
\def\UTA{Department of Physics, University of Texas, Arlington, TX 76019, USA}
\def\Oregon{Center for High Energy Physics, University of Oregon, Eugene, Oregon
97403-1274, USA}
\def\Berkeley{ Department of Physics, University of California, Berkeley, CA 94720, USA}
\def\LBNL{Theoretical Physics Group, Lawrence Berkeley National Laboratory, Berkeley,
CA 94720, USA}
\def\Valencia{IFIC, University of Valencia, Valencia, SPAIN}
\def\Kyushu{Department of Physics, Kyushu University, Fukuoka, JAPAN}
\def\Kansas{Department of Physics and Astronomy, University of Kansas,
  Lawrence,  KS 66045, USA}

\def\Title#1{\begin{center} {\Large #1 } \end{center}}
\def\Author#1{\begin{center}{ \sc #1} \end{center}}

\newcommand\pubblock{\rightline{\begin{tabular}{l} \pubnumber\\
         \pubdate \end{tabular}}}
\newenvironment{Abstract}{\begin{quotation} \begin{center}
                       ABSTRACT
     \end{center}\bigskip  }{\end{quotation}}

\def\beq{\begin{equation}}
\def\eeq#1{\label{#1}\end{equation}}
\def\eeqn{\end{equation}}

\newenvironment{Eqnarray}%
   {\arraycolsep 0.14em\begin{eqnarray}}{\end{eqnarray}}
\def\beqa{\begin{Eqnarray}}
\def\eeqa#1{\label{#1}\end{Eqnarray}}
\def\eeqan{\end{Eqnarray}}
\def\CR{\nonumber \\ }

\def\leqn#1{(\ref{#1})}

\let\bar=\overbar

\def\etal{{\it et al.}}

\def\eg{{\it e.g.}}

\def\lsim{\mathrel{\raise.3ex\hbox{$<$\kern-.75em\lower1ex\hbox{$\sim$}}}}
\def\gsim{\mathrel{\raise.3ex\hbox{$>$\kern-.75em\lower1ex\hbox{$\sim$}}}}

\def\L{{\cal L}}

\def\L{{\cal L}}

\def\half{\frac{1}{2}}

\def\del{\partial}
\def\Dslash{\not{\hbox{\kern-4pt $D$}}}
\def\dslash{\not{\hbox{\kern-2pt $\del$}}}

\def\Dlr{\mathrel{\raise1.5ex\hbox{$\leftrightarrow$\kern-1em\lower1.5ex\hbox{$D$}}}}

\def\ee{e^+e^-}
\def\sstw{\sin^2\theta_w}

\def\mw{m_W}

\def\msb{{\bar{\scriptsize M \kern -1pt S}}}

\def\drb{{\bar{\scriptsize D \kern -1pt R}}}
\def\ELER{e^-_Le^+_R}

\makeatletter
\def\section{\@startsection{section}{0}{\z@}{5.5ex plus .5ex minus
 1.5ex}{2.3ex plus .2ex}{\large\bf}}
\def\subsection{\@startsection{subsection}{1}{\z@}{3.5ex plus .5ex minus
 1.5ex}{1.3ex plus .2ex}{\normalsize\bf}}
\def\subsubsection{\@startsection{subsubsection}{2}{\z@}{-3.5ex plus
-1ex minus  -.2ex}{2.3ex plus .2ex}{\normalsize\sl}}

\renewcommand{\@makecaption}[2]{%
   \vskip 10pt
   \setbox\@tempboxa\hbox{\small #1: #2}
   \ifdim \wd\@tempboxa >\hsize     
       \small #1: #2\par          
     \else                        
       \hbox to\hsize{\hfil\box\@tempboxa\hfil}
   \fi}

\makeatother

\begin{document}
\begin{titlepage}
\pubblock

\vfill
\Title{Tests of the Standard Model  at the}
\Title{International Linear Collider}

\bigskip

\bigskip 

\Author{LCC Physics Working Group}
\vfill

\Author{Keisuke Fujii$^1$, Christophe Grojean$^{2,3}$, Michael
  E. Peskin$^4$  (Conveners); Tim Barklow$^4$, Yaunning Gao$^5$,
  Shinya Kanemura$^6$, Hyungdo Kim$^7$, Jenny List$^2$, Mihoko
  Nojiri$^{1,8}$, Maxim Perelstein$^{9}$, Roman P\"oschl$^{10}$,
  J\"urgen Reuter$^{2}$, Frank Simon$^{11}$, Tomohiko Tanabe$^{12}$,
  James D. Wells$^{13}$, Jaehoon Yu$^{14}$;   Junping Tian$^{12}$,
Taikan Suehara$^{15}$, Marcel Vos$^{16}$, Graham Wilson$^{17}$;
  James Brau$^{18}$, Hitoshi Murayama$^{8,19,20}$ (ex officio)}

\vfill

\begin{Abstract}
We present  an overview of the capabilities that the International Linear
  Collider (ILC) offers for precision measurements that probe the Standard
  Model.  First, we discuss the improvements that the ILC will make in
  precision electroweak observables, both from $W$ boson
  production and radiative return to the $Z$ at 250~GeV in the center
  of mass and from a dedicated GigaZ stage of running at the $Z$
  pole.   We then present new results on precision measurements of
 fermion pair production, including the production of $b$ and $t$
 quarks.   We update the ILC projections for the determination of
 Higgs boson couplings through a Standard Model Effective Field Theory
 fit taking into account the new information on precision electroweak
 constraints.  Finally, we review the capabilities of the ILC to
 measure the Higgs boson self-coupling.
\end{Abstract}
\vfill

\newpage
\phantom{top of page}

\vfill

\noindent { \it
$^1$  \KEK \\ 
$^2$  \DESY\\
$^3$ \Berlin \\ 
$^4$ \SLAC \\ 
$^5$ \Peking\\
$^6$   \Osaka \\ 
$^7$  \SNU \\ 
$^8$ \IPMU \\ 
$^9$ \Cornell \\ 
$^{10}$   \Orsay \\ 
$^{11}$  \Munich \\ 
$^{12}$  \Tokyo\\
$^{13}$ \Michigan \\ 
$^{14}$ \UTA \\ 
$^{15}$   \Kyushu\\ 
$^{16}$  \Valencia \\ 
$^{17}$  \Kansas\\ 
$^{18}$   \Oregon\\ 
$^{19}$  \Berkeley \\ 
$^{20}$  \LBNL\\  }

\vfill

\newpage

\tableofcontents
\end{titlepage}

\newpage

\def\thefootnote{\fnsymbol{footnote}}
\setcounter{footnote}{0}

\section{Introduction}
\label{sec:intro}

Given the central role of the Higgs boson in the Standard Model of particle physics,
the detailed study of the properties of the Higgs boson will be a major goal of
future particle physics experiments. The planned future running of the Large Hadron Collider will 
improve our knowledge of the Higgs boson, as documented in the
report  \cite{Cepeda:2019klc} on the prospects for Higgs studies in its high-luminosity phase.  However,
 a true high-precision understanding of the Higgs boson, capable of discovering 
new physics through the Higgs boson over a wide range of models, has an even more challenging 
requirement: It demands that we push the uncertainties in Higgs boson couplings below the 1\% level~\cite{Dawson:2013bba}.  This  will require studies of Higgs boson production at 
an $\ee$ collider.  A number of $\ee$ Higgs factories have been proposed and are
now in various stages of readiness for construction.

It has recently become clear that the best way to extract the values of the couplings of the 
Higgs boson from experimental observables is to make use of Standard Model Effective Field Theory (SMEFT)~\cite{Durieux:2017rsg,Barklow:2017suo, deBlas:2019rxi,deBlas:2019wgy}. In this method, deviations in the Higgs couplings from the predictions of the Standard Model (SM) are parametrized by the addition to that model of the most general set of dimension-6 gauge-invariant operators.   This is a very general parametrization that can incorporate the effects of almost any type of new physics that can modify the SM at high energies.   The SMEFT method gains its power from unifying constraints on the SM that come from many sources, including not only Higgs measurements but also measurements of precision electroweak observables, 
triple gauge boson couplings, and two-fermion production processes including top quark production.  One of the advantages of $\ee$ colliders is that they offer a large number of well-characterized, independent observables, enough to determine independently
 each coefficient for the full set of  operators that contribute to 
Higgs boson processes. The number of observables becomes even larger, so that the fit is 
actually overconstrained, with the use of polarised beams. Using this method, 
we can extract the Higgs boson couplings from experimental observables in $\ee$ collisions  with no model-dependent assumptions other than the 
validity of the SMEFT itself. 

The goal of this paper is to explain systematically the determination of these SMEFT 
coefficients at the $\ee$ Higgs factory that we consider closest to realization---the International Linear Collider (ILC) in Japan.  We will present estimates for the precision with which all relevant SMEFT parameters will be determined in the proposed ILC program.  As has been explained in previous expositions on the ILC, these estimates come from  full-simulation analyses based on detailed detector models~\cite{Behnke:2013lya,Bambade:2019fyw}.   Thus, we have very high confidence that the precision we claim for these measurements can be realized in practice when  the ILC is constructed.

Our projections for the ILC uncertainties in  precision electroweak observables and our updated projections for Higgs boson couplings are presented in the tables in Appendix~A. 

The outline of this paper is as follows:   In Section 2, we will describe the expected run plan of the ILC.  The minimal plan for the ILC includes running at 250~GeV, 350~GeV, and 500~GeV with polarised beams.   The ILC is also capable of a run at the $Z$ pole (``GigaZ'') with minimal modification.   By extending the length of the linacs, the ILC can run at 1~TeV with the same 
accelerator technology.   The machine parameters for all of these settings have been described previously~\cite{Bambade:2019fyw,Barklow:2015tja}.   In Section 2, we will review the plans
for each stage, giving for each the expected integrated luminosities and calendar durations.

Precision measurements at $\ee$ colliders depend crucially on a precise knowledge of the beam parameters and the detector performance.   Especially for precision electroweak measurements, large event samples are not useful unless one can 
ensure that the experimental systematic errors are sufficiently small. Linear colliders such as the ILC   offer the possibility of longitudinal polarisation both for the electron and positron beams.  We will see that the use of polarisation allows us to design measurements in which the systematic errors on beam parameters are the dominant ones that must be considered.
  In Section 3, we will explain how the rather ambitious goals 
this requires for the systematic errors on beam polarisations and energies will be met.

Following this introduction, we present our survey of ILC physics results.  We have previously presented detailed 
discussions of the measurements of  single-Higgs production and $W$ boson couplings   in  \cite{Bambade:2019fyw}.   Here, we will only give updates to these
measurements.  Our main focus will be on precision electroweak observables and observables of fermion pair production.

We begin in 
Section 4 with a discussion of the $W$ boson mass and width. We  will describe the measurement of the mass of the $W$ boson from kinematic fitting to $\ee\to W^+W^-$ events at 250~GeV.
Section~4 will also  briefly describe the measurement of the $W$ mass that would be possible in a dedicated run at the $WW$ threshold.   

Section~5 gives an introduction to the $Z$ pole observables that we will discuss in this 
report.  The best possible ILC measurements of precision electroweak parameters will be obtained from a dedicated GigaZ run at the $Z$ pole.   However, data at 250~GeV taken as a part of the Higgs boson study will already markedly improve our knowledge of precision electroweak observables beyond what 
is known today.  Section~6  will describe the measurement of $Z$-fermion couplings from the radiative return process 
$\ee\to Z\gamma$ at 250~GeV.   Section~7 will describe the GigaZ program  and the further improvement of precison electroweak measurements that will be possible there.   The 
ILC expectations for precision electroweak measurements are summarized in Table~\ref{tab:PEWresults} in Appendix~A. 

We then turn to the precision measurement of fermion pair production at 250~GeV and at higher
energies.  Section~8 will present the expectations for ILC precision tests of  of $\ee\to f\bar f$ cross sections, including constraints on $s$-channel $Z'$ bosons.   This section will also describe the measurement of four-fermion contact interactions at the ILC.    These contact interactions may, in principle, depend on fermion flavor and helicity.  Current limits from the LHC are given in schemes restricted by model-dependent relations among these couplings.   At an $\ee$ collider with polarised beams, each individual coefficient of a contact interaction can be measured independently.  

 Section~9 will discuss special aspects of the pair-production of bottom and top quarks.   The treatment of these heavy quarks in SMEFT is especially  complicated, requiring 17 independent operators beyond those appropriate for processes with light flavors.  In this section, we will explain how these operator coefficients can be measured in the ILC runs at 250~GeV, 500~GeV, and 1~TeV.   The various coefficients can be completely disentangled in a model-independent
 way~\cite{Durieux:2018tev,Durieux:2019rbz}.  This study is interesting in its own right, because the precision study of the top quark can hold its own clues to possible new physics.  But, also, these results allow a precise determination of the top quark Yukawa coupling in a way that is free of model-dependent assumptions needed for the extraction of this coupling at hadron colliders.

Section~10 will put all of these pieces together and present the expected results from  the global SMEFT fits for Higgs boson couplings that will be possible at the 250~GeV, 500~GeV, and 1~TeV stages of the ILC.   This program is capable of bringing the uncertainty on all major Higgs couplings below the 1\% level. We also discuss the effect on this analysis of higher-precision results from GigaZ.  The ILC expectations for the measurement of 
Higgs boson couplings are summarized in Table~\ref{tab:ILCHiggs}  in Appendix A. 

Section~11 will describe the determination of the Higgs self-coupling at the ILC,  first directly from the measurement of double Higgs boson production at 500~GeV and 1~TeV, and also indirectly by the inclusion of a parameter for the Higgs self-coupling in the SMEFT fit.  We will show that the ILC at 1~TeV will offer multiple determinations of the Higgs self-coupling.  The combination will give a precision better than 10\%, determined in a manner that is free of model-dependent assumptions on the nature of new physics.

\section{ILC accelerator run plan and options}
\label{sec:accel}

To discuss the capabilities of the ILC, it is first necessary to
specify the run plan in terms of energies, integrated luminosities,
and polarisation settings.   The current proposed run plan for the ILC
raises the CM energy in stages, with runs at 250~GeV, 350~GeV, and
500~GeV.   It is also possible to run the ILC at the $Z$ pole  with
minimal modification.  By lengthening the ILC tunnel, improving the
gradient of the superconducting RF cavities, or a combination of
these, it is possible to run the ILC at a CM energy of 1~TeV. All of
these possibilities have been described, together with the necessary
machine parameters, in the previous ILC 
reports~\cite{Bambade:2019fyw, Barklow:2015tja,Adolphsen:2013jya,Adolphsen:2013kya}.

\begin{figure}[tb]
\begin{center}
\includegraphics[width=0.70\hsize]{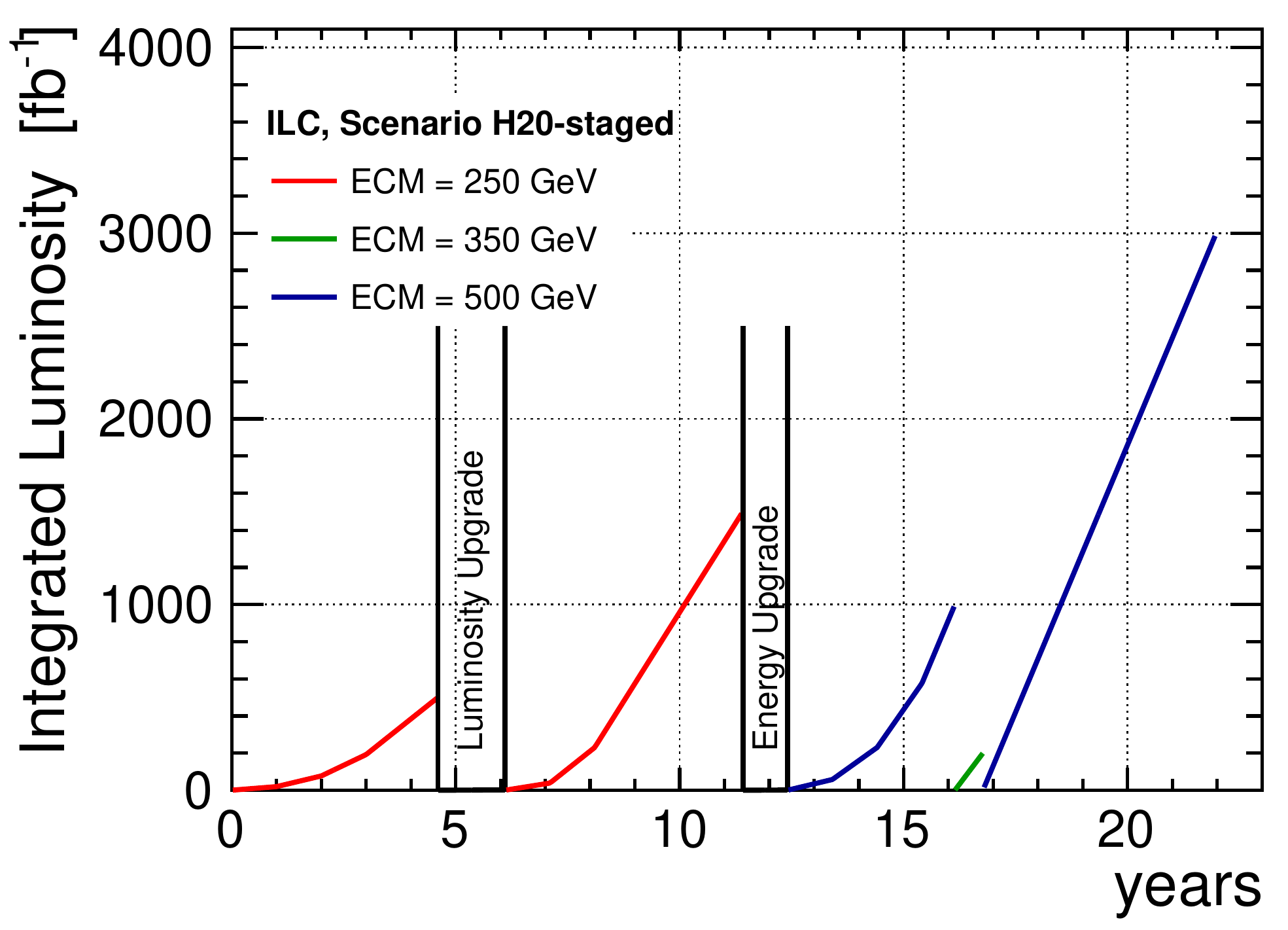}
\end{center}
\caption{The nominal 22-year running program for the staged ILC, starting operation at 250~GeV~\cite{Bambade:2019fyw}. }
\label{fig:H20staged}
\end{figure}

\subsection{Minimal plan}

The currently proposed run plan, in terms of energy and luminosity, is illustrated in
Fig.~\ref{fig:H20staged}~\cite{Bambade:2019fyw}.   
The initial running of the ILC will be at a CM energy of 250~GeV
  with 
bunch trains of 
1312 $e^-$ or $e^+$ bunches per linac pulse, ramping up to
an instantaneous luminosity of   $1.35\times 10^{34}$cm$^{-2}$sec$^{-1}$.
After 6 years, additional RF power will be added, increasing the
number of bunches per linac pulse to 2625 and doubling the
instantaneous luminosity.   This is a relatively inexpensive change,
estimated 
at 8\%
of the initial ILC cost.  It is referred to in the figure as the
``Luminosity Upgrade''.  After reaching a total integrated
luminosity of  2 ab$^{-1}$, the linacs would be lengthened to provide
a CM energy of 500~GeV.  This is referred to in the figure as the ``Energy
Upgrade''.    In fact, if funds are available, most of this upgrade
could be prepared in parallel with physics running at 250~GeV.   The
extended machine would then ramp up to an instantaneous luminosity of 
 $3.6\times 10^{34}$cm$^{-2}$sec$^{-1}$ and acquire  4 ab$^{-1}$ of data, with
 a brief interval of running at  350~GeV to measure the top quark mass
 with high precision.  The luminosity of a linear collider naturally
 rises approximately linearly with CM energy, making it easier to
 acquire larger luminosity samples as the energy is increased.

The ILC is designed to provide significant polarisation for both the
electron and positron beams. We expect $\pm 80\%$ polarisation for the 
electron beam and $\pm 30\%$ polarisation for the positron beam. 
Beam polarisation plays an important role
in the ILC physics, both in producing additional observables with
significant physics information and in controlling systematic errors.
The   importance of polarisation at the ILC is discussed in detail in 
\cite{MoortgatPick:2005cw,Fujii:2018mli}.  Thus, for each operating
energy of the ILC, one must also specify the fraction of time that
will be spent in each of the four possible polarisation states.  Our
baseline
choices are given in Table~\ref{tab:stageswpol}.   Note that the
physics studies at 1~TeV, described below, assumed a positron
polarisation of $\pm 20\%$. 

The full calendar duration of the minimal ILC plan shown in
Fig.~\ref{fig:H20staged}
is 22~years.   However, the plan for the ILC allows additional stages
of running either interleaved with those just described or carried out
after the end of the program.   In this report, we will discuss
results for a GigaZ stage at $Z$ resonance and for an ILC  stage at
1~TeV.  
 The GigaZ program, in particular, could
be carried out within or after the 250~GeV stage  or within the
500~GeV 
stage, whenever its physics results are deemed to be required. 
In the following, we will refer to the stages of the ILC as ILC250,
ILC350, {\it etc.}, following the nomenclature of  Table~\ref{tab:stageswpol}.

\begin{table}[tb]
\begin{center}
\begin{tabular}{lcc|cccc}
  &             & $\int \L$  &   \multicolumn{4}{c}{fraction
 with $\mathrm{sign}(P(e^-),P(e^+))= $ }  \\ 
  & $ E_{CM}$ (GeV) & (fb$^{-1}$) &   $(-+)$  &  $(+-)$  &  $(--)$ &  $(++)$ \\ \hline
  ILC250   & 250 &   2000    &   45\%  &   45\%  &  5\% &  5\% \\
  ILC350   & 350 &  200   &   67.5\% &  22.5\% & 5\%  & 5\% \\
  ILC500  &  500  &   4000    &   40\%  &   40\%  &  10\% &  10\% \\ \hline
  GigaZ   &  91.19 &   100   &     40\%  &   40\%  &  10\% &  10\% \\
 ILC1000   & 1000 &   8000  &     40\%  &   40\%  &  10\% &  10\% \\
 \end{tabular}
 \end{center}
\caption{CM energy, integrated luminosity, and polarisation fractions
  for
 the stages of ILC discussed in this report.  In all cases, the
 magnitude of the $e^-$ polarisation is taken to be 80\% and 
the magnitude of the $e^+$ polarisation is taken to be 30\%, 
except that, at ILC1000,  20\% $e^+$ polarisation was used in the studies quoted.}
\label{tab:stageswpol}
\end{table}

\subsection{GigaZ}
\label{sec:gigazaccel}

Although  a physics run at the $Z$ pole is not part of the minimum 
baseline run plan of the ILC, 
it has always been considered as an important option which should not
be obstructed by the accelerator design.  In particular, the GigaZ operation was
considered 
in the 2015 study by the Joint Working Group on ILC
Beam Parameters~\cite{Barklow:2015tja}.   That group recommended the
following run scenario as the canonical one for physics studies: The 
integrated luminosity should be taken as $100$\,fb$^{-1}$.  Both beams are
assumed to be polarised, with the polarisation fractions 
as in Table~\ref{tab:stageswpol}. 

\begin{table}[t]
\centering
\renewcommand{\arraystretch}{1.10}
\begin{tabular}{l|c c c c|c} 
 \hline
         &  \multicolumn{4}{c|}{$\mathrm{sign}(P(e^-),P(e^+))= $ } & \\
                                 & $(-,+)$       & $(+,-) $      & $(-,-) $      & $ (+,+) $ &  sum  \\
 \hline
 \hline
 luminosity   [fb$^{-1}$]        &    40        &   40        &   10         &    10   &      \\
 $\sigma(P_{e^-},P_{e^+})$ [nb]  &    60.4     &   46.1      &   35.9
                                               &    29.4  &      \\
 $Z$ events [$10^9$]             &    2.4       &   1.8       &   0.36       &    0.29 &  4.9 \\
 hadronic $Z$ events  [$10^9$]   &    1.7       &   1.3       &   0.25       &    0.21 &  3.4 \\
\hline
\end{tabular}
\caption{Integrated luminosities per beam helicity configuration for $Z$ pole running of the ILC, along with the corresponding cross sections and numbers of produced $Z$'s.}
\label{tab:pollumiNZ} 
\end{table}

Table~\ref{tab:pollumiNZ} shows the resulting distribution of the
luminosity onto the four polarisation sign configurations, along with
the corresponding polarised cross 
sections for $|P_{e^-}|=80\%$ and $|P_{e^+}|=30\%$
and the number of produced (hadronic) $Z$ events. 
These have been calculated based on the values for the unpolarised
peak hadronic cross section including QED radiative corrections,
$\sigma= 30$\,nb~\cite{ALEPH:2005ab}, and the left-right asymmetry, 
$A_{LR} = (A_e) = 0.1515$~\cite{Tanabashi:2018oca}. The last 
column gives the total number of (hadronic) $Z$ events summed over 
all data sets. Thus, ``GigaZ'' is actually nearly 5 (3.5) $\times
10^9$ $Z$ events in all (hadronic) decay modes. 

The presence of four data sets of different polarisation signs
 allows a very precise and robust determination of the left-right
 asymmetry of the $Zee$ coupling, as we will 
describe in Section~\ref{sec:gigaz}~\cite{MoortgatPick:2005cw}.

There are different schemes for  implementing the $Z$ pole operation
at the ILC, depending on the machine stage at the time that this run
is scheduled.   The actual running time required to collect the GigaZ
event sample depends on this implementation and can 
range between 1 and 3 years.
None of the possible implementations has been studied at a level of
detail comparable to the ILC baseline. Therefore,  the estimates in
this report  are 
very conservative.  They  are expected to improve 
with further optimisation of the machine design.  

Originally, the implementation of the GigaZ option was studied for the
case of
 the 500\,GeV machine~\cite{ILCpossibilities}.  With that as a
 starting point, the electron linac would be operated at 5+5 =
 10\,Hz,
 alternating between pulses accelerated to $M_Z$/2 for collisions and
 pulses 
accelerated to $150$\,GeV for positron production. Higher luminosities
could be reached by splitting the electron linac into separate halves
devoted to these two purposes.

Recently, the situation was reconsidered assuming that the GigaZ run
would be done after the  first stage of the
ILC at 250\,GeV~\cite{Yokoya:2019rhx}.   Without assuming the installation of any
additional cryogenic power, the electron linac could be operated at
3.7\,Hz + 3.7\,Hz, alternating between acceleration to $M_Z$/2 and to
the nominal $125$\,GeV.  This proposal also takes advantage of a
recent  optimisation of the machine design, allowing for a smaller horizontal 
emittance achieved in the damping rings. 
After a preliminary study of the emittance growth
 along the linacs and of the final focus system, an
 instantaneous luminosity of about $2.1 \times
 10^{33}$cm$^{-2}$s$^{-1}$ 
seems achievable without major modifications.   This luminosity
estimate does not assume the 250~GeV  luminosity upgrade, and so
correspondes to  1315 bunches
per pulse.  If the GigaZ run is done after the  luminosity upgrade, 
to 2625 bunches per pulse, this would
double the GigaZ luminosity to about $4.2 \times 10^{33}$cm$^{-2}$s$^{-1}$.
With the standard ILC assumption of $1.6 \times 10^7$\,s of running per
year, the $100$\,fb$^{-1}$ assumed in the physics studies would correspond to 
about $3.0$ years if done before the luminosity upgrade, or about
$1.5$ years if done
afterward.   If improvement of the precision electroweak measurements
became an important issue, a longer run could be scheduled.   In a 4
year run (as requested for FCC-ee), three times as many $Z$ 
events as assumed above could be collected.

Significant further increases of the luminosity 
could be expected from a more modern design of the 
damping rings, aiming for a smaller longitudinal
 emittance as well as for tighter focusing, and from a better design
 of the Beam Delivery System with a larger aperture of the final focus quadrupoles and
 a larger momentum band width. To put these improvements 
on a solid footing, additional studies are needed. 
Therefore we currently
do not increase the assumed size of the data set beyond 
$100$\,fb$^{-1}$ for physics studies, although the above 
discussion shows that larger data sets could possibly be obtained
 if the physics need arises.

\subsection{1 TeV}
\label{sec:TeVaccel}

The ILC can be upgraded in energy to 1~TeV using current
superconducting RF technology.  Machine parameters for this upgrade
were presented in the ILC TDR~\cite{Adolphsen:2013jya}, Chapter 12.2.
The machine evolution needed is also described in Sec. 2.4.1 of
\cite{Bambade:2019fyw}.   Running of the ILC at 1~TeV looks far enough
into the future that a new generation of accelerator technology will
likely have come into play.   However, many proposals presented to the
2019 European Strategy for Particle Physics are extrapolated over such
long time scale -- for example, FCC presents a 50-year program -- so
our projections for 1~TeV should be taken in the same spirit. 

  The run plan for 1~TeV operation with current technology was
described in Sec.~7 of \cite{Barklow:2015tja}.   There, it is proposed to
acquire a total of  8 ab$^{-1}$ of data.   Both beams are
assumed to be polarised, with polarisations of 80\% and 20\% for the electrons
and positrons, respectively, with polarisation fractions as detailed
in 
Table~\ref{tab:stageswpol}.  Since the
luminosity of a linear collider naturally increases with the
CM energy, the calendar time for this run would be similar to that for
the 4 ab$^{-1}$ run at 500~GeV, that is,  7--8 years. 

The ILC at 1~TeV has interesting capabilities to search for new
color-singlet particles.  It will extend the search reach for pair-production of
dark matter particles, using the mono-photon signature,  and for
electroweakinos and similar particles with compressed spectrum to
interesting
and relevant regions of parameter space.  In this
report, 
however, we will concentrate on the expected results in Higgs boson
physics.

By the end of the 500~GeV ILC program,  we hope that a new high-gradient
accelerator technology will be ready to form the basis of a successor
to the ILC.   Ideas for electron acceleration at a few GeV/m that are
now being investigated would produce an 
electron collider in the same tunnel as the ILC at multi-10~TeV
energies.   We see this as the true long-term future of the ILC laboratory.   However,
the 1~TeV run of the ILC with current technology 
 is something that we can propose now and
investigate with our current analysis tools.

\section{Measurement of polarisation and beam energy at the ILC} 
\label{sec:pol}

Precise knowledge of the electron and positron beam  energy and polarisation is essential
for many measurements at the ILC, and in particular for the Higgs and electroweak program.
For both energy and polarisation, the final values will be obtained by combining measurements from dedicated beam instrumentation with information extracted from the electron-positron collisions themselves. The beam instruments --- polarimeters and energy spectrometers~\cite{Boogert:2009ir} --- provide fast measurements which can resolve the time-dependence of beam parameters during a run, while the collision data provide  the long-term overall scale calibration. In the following, we will summarize the state-of-the-art concepts for polarimetry and beam energy measurement which lead to the estimates of the associated systematic uncertainties used in the remainder of this document.

\subsection{Beam polarisation measurement}
\label{sec:beampol}

The polarisation of the electron and positron beams will be measured by Compton polarimeters located
 about $1.8$\,km before and $100$\,m behind the $e^+e^-$ interaction point (IP). These polarimeters have been designed to measure the ``instantaneous'' longitudinal polarisations at the polarimeter locations with negligible statistical uncertainties and a systematic uncertainty not larger than $\Delta P/P = 0.25\%$~\cite{Vormwald:2015hla, List:2015lsa}. Optionally, the transverse polarisation components could
also  be measured~\cite{Mordechai:2013zwm}.

These polarimeter measurements need to be corrected for the spin transport through the magnets of the Beam Delivery System as well as for the depolarisation in the collisions themselves. These effects have been investigated in detail for ILC500, with the result that 
 they can be controlled to the level of $0.1\%$~\cite{Beckmann:2014mka}, provided that both the relative alignment of the orbit between the $e^+e^-$ IP and the polarimeter locations as well as other beam parameters (energy, intensity, emittance, $\beta$ function at IP) are sufficiently well known.  With such corrections, the luminosity-weighted long-term average of the polarisation at the $e^+e^-$ IP can then in principle be 
derived from the polarimeter measurements.

 The polarisations of the electron and positron beams can also be measured from the observed $\ee$ cross sections.  What makes this strategy effective is that, at energies above the $Z$ pole, well-understood processes such as forward $W$ pair production have cross sections with  strong polarisation-dependence. The estimate of polarisation from these
 measurements does not rely on the modeling of the accelerator parameters, but high precision is obtained only by integrating over long periods, which washes out time-dependent variations. 
Therefore, the extraction of the polarisation from collision data and the measurements by the
polarimeter complement one another and will be combined to achieve
 the  ultimate level of  precision.

The extraction of the luminosity-weighted long-term average polarisations from collision data has been the subject of many studies. The latest status can be found in~\cite{ThesisRobert}.  Observations to note are:
\begin{itemize}

\item It is important to have four independent beam settings with positive and negative values for both beam polarisations $P_{e^-}$ and $P_{e^+}$.  It is not sufficient to assume that the absolute value of the polarisation stays the same when the polarisation is reversed, as in the modified Blondel 
scheme~\cite{Blondel:1987wr,Monig:2000bn}.  Instead,  it is necessary to correct the results from the Blondel scheme based 
on the polarimeter measurements, and it is this effect that actually limits the  final precision~\cite{ThesisRobert}.

\item When the  total and differential cross sections for  2- and 4-fermion processes are combined to extract the four polarisation parameters, these parameters  can be determined to a few parts in $10^4$~\cite{ThesisRobert}, provided
that the  efficiencies and purities of the event selections can be controlled at the per mille level. This justifies the estimates used in this paper that the relative systematic errors on left-right asymmetries due to the uncertainty of the beam polarisations is about $3 \times 10^{-4}$ at $\sqrt{s}=250$\,GeV. At the $Z$ pole, no 4-fermion processes
will be available; therefore we expect  larger systematic uncertainties of $5 \times 10^{-4}$.
\item We are currently developing a superior method for the estimate of polarisation uncertainties:  the inclusion of the polarisation values  as nuisance parameters in the extraction of the main observables in 
$\ee$ cross section measurements.  This has been done routinely in the projections for triple gauge coupling precisions~\cite{Marchesini:2011aka, ThesisRobert}, and has recently been started also for the extraction of other electroweak parameters~\cite{ThesisRobert}. In a simultaneous extraction of 
the beam polarisations at the ILC250, the total unpolarised cross sections and the left-right asymmetries of various
 2- and 4-fermion processes using  a fit to the total and differential cross section measurements of these 
processes, precisions of $4.3 \times 10^{-4}$ and $5 \times 10^{-4}$ have been obtained for $A_{LR}(e^+e^- \to q \bar{q})$ and $A_{LR}(e^+e^- \to l^+ l^-)$, respectively.  These numbers, which  combine the statistical uncertainty and the uncertainty due to the finite knowledge of the polarisation,  are even better than the results we present later in this document.
 However, since this study is not yet based on full simulation,
 we only take it here as additional support that the precisions defined in the previous item can actually be achieved. 

\item The availability of positron polarisation is very important to achieve a small systematic uncertainty on the  polarisation, since it introduces redundancy which overconstrains the system. Without positron polarisation, the uncertainties on $A_{LR}$ from the global fit discussed in the previous item would be larger by a factor of $10$.
\end{itemize}

\subsection{Beam energy measurement}
\label{sec:beamenergy}

For each beam, energy spectrometers~\cite{Boogert:2009ir} will measure the beam energy before and after the collision point to a precision of $\delta E_b / E_b = 10^{-4}$. The upstream and downstream spectrometers are based on complementary
approaches and have been designed to achieve the target precision over the full range of possible ILC beam energies from $45.6$ to $500$\,GeV.

These measurements can be augmented by exploiting the collision data themselves.
In particular, in  $e^+e^- \to \mu^+\mu^- \gamma$ events, the transverse momenta
and angles of the muons are precisely measured using  the tracking system of the detectors.  
Then, from energy and momentum conservation, the 
center-of-mass energy of the collision can be extracted without the need to detect the photon.
This method can reach statistical precisions of $\delta E_b / E_b = 10^{-5}$ or better at all center-of-mass energies. The dominant systematic uncertainty is the  momentum scale of the detector~\cite{bib:graham_ecfa2013}. 

The detector's momentum scale can be calibrated using the 
$J/\Psi$, whose mass is known to $1.9 \times 10^{-6}$~\cite{Tanabashi:2018oca}. Based on a run at the $Z$ pole providing $10^{9}$ hadronic $Z$ events, the statistical uncertainty on the reconstructed mass peak position in di-muons will be smaller than $2 \times 10^{-6}$~\cite{bib:graham_awlc2014}. This means that the absolute momentum scale can be determined sufficiently well to match the requirements of the method described in the previous paragraph and produce an absolute 
$\delta E_b / E_b$ of $ 10^{-5}$. 

The point-to-point energy uncertainty, \eg,  during a resonance or 
threshold scan, will reach similar precisions.
    While the absolute energy scale does not affect the measurements 
of the point-to-point variations, the statistics of each data set during
    a scan will usually be smaller than in the main (peak) sample.  
Thus, the in-situ methods will typically be statistically limited: for a 10 times
    smaller sample, the statistical precision will be a factor 3 worse.
 On the other hand the beam energy spectrometer precision given above is by far
    dominated by the absolute calibration of the beam position 
monitors. Relative changes in the beam position can be measured much more precisely.
    Therefore the combination of in-situ methods and the energy spectrometers will allow a determination of the energy of smaller data sets with precision equal to that of
 the absolute energy of a main data set.

\subsection{Luminosity measurement}
\label{sec:beamlumi}

At all lepton colliders, the luminosity spectrum is broadened by initial-state radiation.  At linear colliders (and high-luminosity circular colliders) there is additional radiation due to the beam-beam interaction (``beamstrahlung'').  
The exact shape of the distribution of the luminosity as a function of the event-by-event CM energy is called the luminosity spectrum. This spectrum has a peak near the nominal CM  energy with spread given by the intrinsic energy spread of the beams, which is estimated to be $10^{-3}$~\cite{Adolphsen:2013kya}, and a long tail to lower  CM energies due to beamstrahlung.  The average energy loss in this tail is a few percent at ILC250 but becomes increasingly important at higher CM energies.

Beamstrahlung depends  on the 
instantaneous machine parameters and thus must be directly measured.  In the study ~\cite{Poss:2013oea} for CLIC at $\sqrt{s} =$~3~TeV, the luminosity spectrum was reconstructed
from radiative Bhabha events bin-by-bin with a maximum  error of 5\% over the whole energy range, leading to a residual systematic effect on energy and mass observables of a few 10's of 
MeV. At the ILC, with much less beamstrahlung, 
the precision is expected to improve to the percent level.
 As for the polarisation and beam energy measurements, a long-term determination of the luminosity spectrum from physics events will be combined with fast extractions of beam parameters from the pattern of $e^+e^-$ pairs and photons in the very forward calorimeters BeamCal and GamCal~\cite{Grah:2008zz}, which can be performed every few bunch crossings. An example of propagating
the results from~\cite{Grah:2008zz} to a full physics analysis can be found in~\cite{Habermehl:2018yul}.  In this example, 
the effect on signal and background predictions is found to be at the few per mille level even when only using the ``online'', BeamCal-based method and not the full Bhabha analysis.

The absolute luminosity above $80\%$ of the nominal center-of-mass energy can be determined from low-angle Bhabha scattering measured in the dedicated forward calorimeters of the ILC detectors to a precision of a few per  mille~\cite{Bozovic-Jelisavcic:2013aca}.

\section{Precision W measurements at 250 GeV}  
\label{sec:Wmass}

Two of the main experimental observables of interest for
precision tests of the SM in the $W$ boson sector 
are the $W$ mass, $\mw$, and the $W$ width, $\Gamma_W$.
The ILC program with polarised beams and all standard 
stages of the machine is well suited to such measurements, and 
especially at $\sqrt{s}=250$~GeV where data can be collected 
synergistically with Higgs boson related studies.

\subsection{Measurement of $\mw$}

The $W$ mass has been a prime target for the ILC and has been 
understood to be very tractable based on extrapolations of measurements 
from LEP2 both well above $W$-pair threshold, and at $W$-pair threshold. 
Prior prospects for such measurements are summarized in Tables 1-9 
and 1-10 in~\cite{Baak:2013fwa}.

\begin{table}[t]
\begin{center}
\begin{tabular}{lccc}
   $\sigma_{M}$ (GeV)     &  $\Delta \mw$ (MeV)    &  $\Delta \Gamma^{a}_{W}$ (MeV)  &  $\Delta \Gamma^{b}_{W}
$ (MeV) \\ \hline
1.0 & 0.67 & 1.3 & 2.0 \\
2.0 & 0.98 & 1.7 & 2.7 \\
2.5 & 1.1  & 2.0 & 3.2 \\
3.0 & 1.3  & 2.3 & 3.7 \\
4.0 & 1.6  & 2.8 & 5.0 \\ \hline
\end{tabular}
\end{center}
\caption{Statistical uncertainties for 
$\mw$ and $\Gamma_{W}$ expected for a sample of 
 $10^{7}$ reconstructed $W$ bosons at the ILC250. These are estimated 
from a simple parametric fit of the Breit-Wigner 
lineshape convolved with a range of constant Gaussian 
experimental mass resolutions, $\sigma_M$, ranging from 1 to 4 GeV. 
The $\mw$ uncertainty is evaluated with a one parameter fit with 
the width and mass resolution fixed. The corresponding uncertainties on 
the W width are evaluated either with the mass resolution fixed and 
known perfectly from a 
two parameter fit ($\Gamma^{a}_{W}$), or more realistically, 
from a three parameter fit ($\Gamma^{b}_{W}$) that also fits for 
the mass resolution.}
\label{tab:Wmass}  
\end{table}

Measurements from LEP2, the Tevatron, and ATLAS of $\mw$ have led to 
today's precision in the PDG of 12~MeV, with the best single 
experiment measurement having a quoted precision of 18~MeV.
Further improvements from long-existing hadron collider data sets 
at the Tevatron and LHC are possible, and it was suggested in 
\cite{Baak:2013fwa} that the LHC could eventually 
improve the uncertainty on the $W$ mass to 5~MeV. But, given the predominant 
systematic uncertainties, this goal now looks very challenging. 

It is then interesting to study the challenges to a high-precision 
measurement of $\mw$ at lepton colliders.  Data sets 
at LEP2 totalled 0.7 $\rm{fb}^{-1}$ per experiment, leading to 
statistically limited measurements. The ILC250  is
expected
to produce a 
much larger data  set of 2 ab$^{-1}$, 
with polarised beams.   This data set will provide a sample of more
than $10^7$ reconstructed $W$ bosons.   To demonstrate the statistical
power of such a sample, we show in Table~\ref{tab:Wmass} the
expected statistical uncertainties on $m_W$ and $\Gamma_W$ that would
result from fits to the measured $W$ boson  invariant mass distribution.
For a  typical mass resolution of 2.5~GeV,  $10^7$ $W$ bosons would yield a
 statistical uncertainty 
on $\mw$ of  1.1~MeV.  Similarly, 
fitting the mass, width, and a Gaussian experimental 
mass resolution, the total width could be determined from the 
lineshape with a statistical uncertainty of 3.2~MeV.  Thus, the
measurements of these quantities at the ILC250 will already  reach the regime in
which systematic errors dominate.  We will discuss the expected
systematic errors for each proposed method in the remainder of this
section.  

The $W$ boson total width can also be determined  by the
measurement of the $W$ leptonic branching fractions, since the
absolute leptonic partial widths, including new physics contributions,
can be  predicted from precision electroweak observables. We will discuss
the measurement of $BR(W\to \ell \nu)$ in Sec.~\ref{sec:WBR}.  

There are a number of  promising approaches to measure
the $W$ mass at an $\ee$ collider such as ILC.
Given the data sets that represent more than three orders of 
magnitude increase in statistics beyond LEP2, it is appropriate to also 
consider $W$ mass measurement methods that may have better 
systematic behavior in this high statistics regime.
The various methods for $\mw$ measurement are as follows:

\begin{enumerate}
\item\label{item:kf} {\bf Constrained reconstruction.}
Kinematically-constrained reconstruction of $\mathrm{W}^+\mathrm{W}^-$ 
using constraints from {\it four-momentum conservation} and optionally 
mass-equality, as was done at LEP2.

\item\label{item:mhad} {\bf Hadronic mass.} 
Direct measurement of the {\it hadronic mass}. This can be applied 
particularly to single-$W$ events decaying hadronically 
or to the hadronic system in 
semi-leptonic $\mathrm{W}^+\mathrm{W}^-$ events. 
This method does not rely directly on knowledge of the beam energy
or its distribution.

\item\label{item:endpoint} {\bf Lepton endpoints.}
The two-body decay of each $W$ leads to endpoints 
in the lepton energy spectrum at 
\beq
E_{\ell } = E_{\rm{b}} (1 \pm \beta)/2 \ , 
\eeqn
where  $\beta$ is the $W$ velocity. These can be used to infer $\mw$. 
The endpoints correspond to leptons parallel and anti-parallel 
to the $W$ flight direction
This technique can be used for both semi-leptonic and fully-leptonic 
$WW$ events with at least one prompt electron or muon.

\item\label{item:pm} {\bf Di-lepton pseudo-mass.}
In $WW$ to dilepton events, with electrons or muons, 
one has six unknown quantities, namely, the three-momenta of each neutrino.
Assuming four-momentum conservation and equality of the two $W$ masses, 
one has five constraints. By assuming that both neutrinos are in the 
same plane as the charged leptons, the kinematics can be solved to 
yield two ``pseudo-mass'' solutions that are sensitive to the 
true $W$ mass. This technique was discussed in  Appendix B
  of~\cite{Hagiwara:1986vm}  and 
used along with the lepton endpoints by the OPAL experiment at
LEP2~\cite{OPAL-mW-lvlv}.

\item\label{item:threshold} {\bf Polarised Threshold Scan.}  
Measurement of the $\mathrm{W}^+\mathrm{W}^-$ cross-section near 
threshold with longitudinally 
polarised beams is discussed in~\cite{Wilson:2016hne} and 
references therein. 
The ability to ``turn-on'' and ``turn-off'' the signal with polarised 
beams, a capability unique to ILC,  allows a precise in-situ measurement of the background.
\end{enumerate}

Methods~\ref{item:kf},\ref{item:mhad},\ref{item:endpoint},\ref{item:pm} 
can all exploit the standard ILC program at 250~GeV and above. 
Method~\ref{item:threshold} needs 
dedicated running near $\sqrt{s}=161$~GeV.
Methods for measuring the $W$ mass in $\rm{e}^+\rm{e}^-$ colliders were 
explored extensively in the LEP2 era, see~\cite{Kunszt:1996km,Stirling:1995xp} and 
references therein. 

For ILC-sized data sets, the constrained reconstruction 
approach (method~\ref{item:kf})
may need to be restricted to semi-leptonic events in order to 
avoid the final-state interaction issues 
that beset the fully hadronic channel. With the 
large data-sets of $WW$ events expected above threshold, the expectation 
is that this measurement will be systematics limited. 
With much improved detectors compared to LEP2 and with much better 
lepton and jet energy resolution, it is expected 
that uncertainties in the few MeV level can be targeted. Table 1-9 
in~\cite{Baak:2013fwa} estimates an uncertainty of 2.8~MeV 
at $\sqrt{s}=250$~GeV based on extrapolating LEP2 methods using only 
the semi-leptonic channels with electrons or muons.

Method~\ref{item:mhad} is based purely on the hadronic mass 
and was not used explicitly at LEP2. 
With the increased 
cross-section for singly-resonant events 
($\ee \rightarrow \rm{W} \rm{e} \nu$) at higher $\sqrt{s}$, 
the excellent resolution for particles in jets expected from 
particle-flow detectors, and the availability of control channels 
with hadronic decays of the Z, an opportunity exists to make a 
competitive measurement also using this method. 
However the demands on the effective jet energy scale calibration are 
very challenging. It was estimated 
(Table 1-10 in ~\cite{Baak:2013fwa}) that a $\mw$ 
uncertainty of 3.7~MeV could be reached.  This would be dominated 
by the hadronic energy scale systematic uncertainty.

The endpoints method~\ref{item:endpoint} was only used for 
fully leptonic events at LEP2. It has the inherent advantage that 
the systematic uncertainties are dominated simply by 
the uncertainties on the lepton energy scale 
and the beam energy, given that one can express $\mw$ in terms of the 
endpoints as follows:
\beq
\mw^2 = 4 E_l ( E_{\rm{b}} - E_l )  \ .
\eeqn
It may be worth considering this as a complementary
method also for semi-leptonic events, that is of course correlated with 
the constrained reconstruction method.

The pseudo-mass method and the endpoints method were applied to the 
fully leptonic channel in~\cite{OPAL-mW-lvlv}. Very little 
correlation (+11\%) was found between the two methods, indicating 
that the two  methods can be independently effective and can be 
combined.  The OPAL result achieved a statistical 
uncertainty of 390 MeV on $\mw$ using 0.7 fb$^{-1}$ of data.
The lepton energy resolution for ILC detectors is about 0.15\% 
based on momentum measurements; this is much better than 
the 3\% energy (for electrons) and 8\% momentum (for muons) 
resolutions at OPAL. 
Assuming a factor of two improvement for ILC detectors,  
(note that resolutions much less than $\Gamma_W$ are not necessary), and the 
statistics of the 2 ab$^{-1}$ data set at ILC250.  we project 
a statistical uncertainty on $\mw$ of around 3.6~MeV.
This looks very promising, since the experimental 
systematic uncertainties are very straightforward;  more detailed 
studies should be pursued. With the standard 10 ppm uncertainty 
on center-of-mass energy and detector momentum scale, this approach 
promises to be very fruitful with the full ILC program.

Method~\ref{item:threshold} needs dedicated running near 
$\sqrt{s}=161$~GeV. This is now feasible for 
the ILC machine.  The expected integrated luminosity is
about 125 fb$^{-1}$/year if the run is done after the Luminosity
Upgrade  in Fig.~\ref{fig:H20staged}.  The use of a threshold scan with polarised electron and 
positron beams to yield a precision measurement of $\mw$ at ILC was
studied in ~\cite{Wilson:2016hne}.
One of the potentially dominant systematic uncertainties,
the background determination, is under very good experimental control 
because of the polarised beams. This is difficult to achieve with 
an unpolarised collider. Errors at the few MeV level can be envisaged. 
With 100~fb$^{-1}$, 
and polarisation values of (90\%, 60\%), 
the estimated uncertainty is 
\beq
 \Delta \mw (\rm{MeV}) = 2.4 \: \rm{(stat)} \oplus 3.1 \: (syst) 
\oplus 0.8 \: (\sqrt{s}) \oplus \rm{theory}  \ , 
\eeqn
with these values added in quadrature,
amounting to an experimental uncertainty of 3.9~MeV. With 
standard ILC polarisation values of 80\% and 30\% the 
estimated precision is 6.1 MeV.
Eventual experimental precision approaching 2 MeV from 
this approach can be considered at ILC if one is able to 
dedicate 500 fb$^{-1}$ to such a 
measurement, and the physics perspective of the day demands it. 
There are excellent prospects for very competitive ILC 
measurements of the $W$ mass from the four other methods using data 
collected above the production threshold for Higgs bosons, and so 
it would seem premature to make exclusive 
running at $W$-pair threshold a requirement for the ILC run plan. 
Nevertheless, given the complementary nature of a threshold-based 
measurement it would seem prudent to retain accelerator 
compatibility with such a scenario.

Given that the leading experimental 
    systematic uncertainties for the different methods are 
    reasonably complementary, it is expected that, 
    with the combination of these five different methods, 
    ILC will be able to measure $\mw$ to at least 2.5~MeV. 
    This uncertainty can potentially already be reached with 
    data-taking at the  ILC250.

\subsection{Measurement of $W$ branching fractions}
\label{sec:WBR}

With the large data sets envisaged at ILC250, one can 
also target much improved measurements of 
the $WW$ production cross section and the individual W decay branching 
fractions. This would use the ten different final state cross sections
available from $WW$ production:  
the six $WW$ final states associated with fully leptonic 
final states with two charged leptons (dielectrons, dimuons, 
ditaus, electron-muon, electron-tau and muon-tau), the three 
semileptonic $WW$ 
final states, one for 
each lepton flavor, and the fully hadronic $WW$ final state.  This 
follows the methodology used at  
LEP2~\cite{A-WWBR, D-WWBR, L-WWBR, O-WWBR}.

The ten measured event yields can be fitted for 
an overall $WW$ cross section, $\sigma_{WW}$, and the three individual 
leptonic branching fractions, $B_{e}$, $B_{\mu}$ and $B_{\tau}$ 
with the overall constraint that
\beq
B_{\rm{had}} = 1 - B_{e} - B_{\mu}  - B_{\tau} \ , 
\eeqn
taking into account background contributions in each channel.
With ten channels and four fit parameters, there is some 
redundancy in the fitting procedure.  This means that the parameters can 
be determined well even if the more challenging channels, 
namely the fully hadronic, the semileptonic with a tau,  
and the di-tau channel would end up being systematically limited.
At LEP2, the signal process was modelled simply through the three
dominant, doubly resonant Feynman diagrams (so called CC03 process), 
while other diagrams and their interferences resulting in the 
same four fermion final state, such as those for 
$ZZ$ or $We\nu$,  were treated as background. 
These ``4f-CC03'' corrections were typically about 10\% depending on 
final state.  A complete calculation of $\ee\to 4f$
final states would need to be used in the high statistics regime.

We have looked into the feasibility of this method for ILC250, 
building on LEP2 studies 
at $\sqrt{s}\approx 200$~GeV, by putting together 
a fit ansatz that  assumes 
identical efficiencies and mis-classification 
probabilities for all ten $WW$ channels~\cite{O-WWBR}. 
For the purpose of making an estimate for
this report, we concentrate on the impact of a single subsample of the
data.  The actual analysis at the ILC will be based on global fit to
the 
results  from 
all polarisation modes, along the lines described in
Sec.~\ref{sec:pol}. 

Of the total 2 ab$^{-1}$ to be collected at
ILC250,  
0.9 ab$^{-1}$  is to be collected 
with $\ELER$ enhanced collisions.  These benefit from a $WW$ cross section 
enhancement over unpolarised beams of a factor of 2.32 for $-80\%, +30\% $
beam polarisations. 
The estimated background per selection channel 
depends on the polarisation asymmetry of the different backgrounds
and is estimated to be about +29\% for the important 
two-fermion background from hadronic events. 
Taking this effect that leads to an increased background, 
and the decreased background estimated from $1/s$ scaling, 
we find that the unchanged 
OPAL background estimate is a good first estimate, and believe 
that this is a reasonably conservative estimate.   We have based our 
estimates of statistical errors on the size of this subsample. 
We assume  that the other 55\% of the data set with 
the less favorable beam polarisation configurations is used to measure 
and test the background modeling and have neglected it for now in 
estimating statistical sensitivity.

We also include the 6\% reduction in unpolarised cross section 
at $\sqrt{s}=250$~GeV. Given that ILC detectors will have 
much improved forward hermeticity, jet and lepton 
energy resolutions, vertexing, and 
electron, muon, and tau identification, it is very 
reasonable to expect that the efficiency and background 
performance would be much better.
One effect that is more difficult at higher $\sqrt{s}$ 
is a more forward polar angle distribution of the W decay products. 
We find that 94.7\% of leptons in semi-leptonic events have a polar angle 
satisfying, $|\cos{\theta}| < 0.975$, whereas at $\sqrt{s}=200$~GeV, 
the corresponding fraction is 96.7\%.

It is straightforward to estimate statistical uncertainties 
and we have done so for a number of scenarios. For systematic 
uncertainties, there are five that come to mind:
\begin{itemize}
\item absolute integrated luminosity: The precision is likely limited
  to about 0.1\%; however, to a great extent, this value cancels out
  of the determination of branching ratios.

\item lepton efficiencies: This can be measured with high precision using 
control samples of di-leptons as was done for precise $Z$ lineshape 
measurements preferably using a tag-and-probe method. The key element 
is efficiency within the geometrical acceptance. With control samples 
totalling $10^{7}$ leptons, statistical uncertainties
of $3\times 10^{-5}$ can be targeted assuming highly efficient lepton 
identification.

\item hadronic system modeling: Uncertainties of order 0.03\% seem 
feasible based on LEP1 hadronic $Z$ studies targeted at estimating the 
hadronic efficiency/acceptance.

\item fake $\tau$ candidates from hadronic events:
One needs to be able to model the rate of isolated tracks from 
hadronic systems that can fake tau candidates. 
This should be easier to reduce than at LEP2 given the excellent 
vertexing performance envisaged.

\item background estimation:  This will be  controlled with the less 
signal-favorable beam polarisation configurations.
\end{itemize}

\begin{table}[t]
\begin{center}
\begin{tabular}{lccccc}
  Event selections       &   $B_e$    &  $B_{\mu}$  &  $B_{\tau}$  & $R_{\mu}$ & $R_{\tau}$ \\ \hline
  All 10          &   4.2   &    4.1   &    5.2             &  6.1  &  7.5   \\
  9 (not fully-hadronic)      &    5.9    &     5.7    &   6.4  &  6.1  &  7.5   \\
  9 (not tau-semileptonic)     &    4.6   &     4.6    &   7.8 &  6.1  & 10.8   \\
  8 (not f-h and not $\tau$-semileptonic) & 8.3   &   8.4  &   7.8 &  6.1  & 12.8   \\
  7 (not f-h and not $\tau$-sl and not di-$\tau$) & 9.0  &  9.1  &10.6 &  6.1  & 16.7   \\ \hline
\end{tabular}
\end{center}
\caption{Statistical uncertainties, expressed as relative errors in units of $10^{-4}$ for 
the leptonic branching fractions  of the $W$ boson ($B_e$, $B_{\mu}$
and $B_{\tau}$)
and the ratios of branching fractions $R_\mu = B_\mu/B_e$, $R_\tau =
B_\tau/B_e$.   
The lines of the table refer to different choices of 
the included event selections. The values  assume ILC measurements 
at $\sqrt{s}=250$~GeV using the 45\% of the 2 ab$^{-1}$ integrated 
luminosity with enhanced $\ELER$ collisions, with the same 
efficiencies and the same background 
cross sections as in the OPAL measurement~\cite{O-WWBR}.
The uncertainties given for $R_\mu$, $R_\tau$ are from 
a separate fit using the ($B_e$, $R_{\mu}$ and $R_{\tau}$) parametrization.}
\label{tab:WWBRs}  
\end{table}

In Table~\ref{tab:WWBRs} we show the expected absolute 
statistical uncertainties for two different parameterizations, one based on 
the three leptonic branching fractions, ($B_e$, $B_{\mu}$ and $B_{\tau}$) 
and one based on $B_{e}$ and the ratios $B_{\mu}/B_e$ and 
$B_{\tau}/B_e$. Five different configurations of included event 
selections are considered, indicating a reasonable degree of robustness.
The fits also fit for the cross section but the absolute 
value is likely to be systematics limited.
It can be seen that fractional statistical uncertainties
on $B_e$ below 0.1\% and as low as 0.04\% can be envisaged. 
The fits do not assume lepton universality. 
The data set considered consists of 29.7 million 
$WW$ candidates. The efficiency systematics seem not to be limiting. 
The main systematic issue is likely to be the background 
estimation that should be facilitated with the various polarised 
data sets. The event selection purity 
will likely need to be tightened to reduce systematics from backgrounds, 
but the current statistical estimates should be a reasonable 
starting point.

\section{Precision electroweak observables}
\label{sec:PEW}

Electroweak precision observables measured at LEP and SLC at
the $Z$ pole continue to provide  the backbone of the interpretation of
measurements in the electroweak sector. A comprehensive overview of
these measurements  is
given in~\cite{ALEPH:2005ab}.  In the next few sections, we will explain how 
the ILC will improve on these measurements.

To set up the discussion to follow,  we now define the basic precision
observables.  For simplicity, we express the observables here in terms
of tree-level formulae and describe each observable as having an
independent 
measurement. In practice, the values of observables and the beam
properties will be combined in a global fit, as described in the third
bullet  of  Sec.~\ref{sec:beampol}.

For a given quark or lepton flavor $f$, let $g_{Lf}$, $g_{Rf}$ be the 
helicity-dependent
$Zff$ couplings.  Then the quantities, for quarks $q$,
\beq
        R_q =   {\Gamma(Z\to q\bar q)\over \Gamma(Z\to
        \mbox{hadrons})} \ , 
\eeq{Rfdef}
and, for leptons $\ell  = e, \mu, \tau$, 
\beq
      1/R_\ell =   {\Gamma(Z\to \ell^+\ell^-)\over \Gamma(Z\to
        \mbox{hadrons})} \ , 
\eeq{Rfdef2}
are given,  at the tree level, by 
\beq
    R_q\ ,\  1/R_\ell \propto    (g_{Lf}^2 + g_{Rf}^2) \ , 
\eeq{Rfromg}
and the $Z$ decay polarisation  asymmetries are given by 
\beq
        A_f =  {g_{Lf}^2 - g_{Rf}^2 \over   g_{Lf}^2 + g_{Rf}^2}\ .
\eeq{Affromg}
It is useful to define the value of $\sstw$ governing the $Z$
couplings from the electron asymmetry as ``$\sin^2\theta_{eff}$'' given by
the formula
\beq
    A_e =  { (\half - \sin^2\theta_{eff})^2  - (
    \sin^2\theta_{eff})^2\over  (\half - \sin^2\theta_{eff})^2  + (
    \sin^2\theta_{eff})^2}
        \approx   8 \bigl({1\over 4} - \sin^2\theta_{eff}\bigr) \ .
\eeq{sstweffdef}
It is this value of $\sstw$ that enters the $Zh$ and $WW$ pair
production cross sections that are most important in determining the
Higgs boson couplings.  

Loop corrections to the SM predictions for $Z$ observables 
 given in terms of $\sin^2\theta_{eff}$ are  at the
 parts per mille level.  Thus, it is  accurate  to quote projections
 for the precision of future experiments
 from tree-level formulae involving $\sin^2\theta_{eff}$.  Of course, actually
 extracting $Z$ couplings from cross section measurements 
 at the $10^{-4}$ level of precision
 requires that the SM contributions to these cross sections be known
 to comparable accuracy.   The nontrivial requirements for theory are
 reviewed in~\cite{Blondel:2019qlh}.

Often, the leptonic asymmetries $A_e$, $A_\mu$, and $A_\tau$ are
combined to give a  composite leptonic asymmetry.  Here, we will
distinguish
these three quantities and discuss tests of models that allow small
differences in the $Z$ couplings to $e$, $\mu$, and $\tau$. 

At a polarised $\ee$ collider, $A_e$ is given by  the left-right
asymmetry
in the total rate for $Z$ production, 
\beq
  A_e =  A_{LR}\equiv {\sigma_L -
     \sigma_R
 \over (\sigma_L + \sigma_R)}  \ ,
\eeq{ALRmeas}
where $\sigma_L$ and $\sigma_R$ are the cross section for 100\%
polarised $e^-_Le^+_R$ and $e^-_Re^+_L$ initial states.   For beams
not perfectly polarised, the effective left-handed polarisation of the
initial state is
given by 
\beq
P_{eff} = (P_{e^-} - P_{e^+})/(1 -  P_{e^-} P_{e^+}) \ , 
\eeq{Peff}
and the measured asymmetry is proportional to $P_{eff}$. 
 The determination of the
quantity $A_e$ then
requires  only an excellent knowledge of the polarisation and knowledge
that the acceptance in the decay modes studied  does not change when
the polarisation is flipped.  Essentially, the entire statistics of
$Z$ production can contribute to the measurement.   We find that the
dominant systematic error is that on the value of the polarisation.
We have discussed how this systematic is controlled in
Sec.~\ref{sec:beampol}. 

For other asymmetries, beam polarisation can also play a role.  These
quantities
are measured from the left-right forward-backward asymmetry
\beq
   A_{FB,LR}^f \equiv {(\sigma_F - \sigma_B)_L -
       (\sigma_F - \sigma_B)_R
 \over (\sigma_F + \sigma_B)_L +
       (\sigma_F + \sigma_B)_R}   \ ,
\eeq{Afmeas}
where, again, L and R refer to states of 100\% polarisation.
At the tree 
level, 
\beq
   A_{FB,LR}^f = {3\over 4} A_f  \ . 
\eeqn
At an unpolarised collider, the values of the $A_f$ are obtained from 
quantities such as the unpolarised forward-backward asymmetries, 
\beq
   A_{FB}^f \equiv {(\sigma_F - \sigma_B)
 \over (\sigma_F + \sigma_B)}   \ .
\eeq{AFBmeas}
At the tree 
level, 
\beq
   A_{FB}^f = {3\over 4} A_e A_f  \ , 
\eeqn
so there is some sacrifice of statistics to achieve the same level of 
precision.   (The determination of $A_\tau$ is a special case, to be
discussed
below.)
For some purposes, for example, to test lepton universality, we wish
to know the ratio of $A_f$ 
to the precisely determined value of $A_e$.  In such ratios of 
polarisation asymmetries measured in the same run, the
systematic uncertainty on the polarisation cancels out. 
 
The uncertainties from acceptance and particle identification largely
cancel out of the $A_f$ measurements, but in the measurements of
$R_f$ they are the major source of systematic error.   In the LEP
experiments, the measurements of the rates of $Z$ decay to $b\bar b$
and $c\bar c$ were mainly done with single-tag methods that required a
``dilution factor'' correction with a large QCD uncertainty.  At the
ILC, the efficiencies for $b$ and $c$ identification
and also the statistics to determine these efficiences precisely, will
be much higher. 
The absolute tagging efficiences can
be measured from $\ee\to f\bar f$ events, using a probe and tag
method.
We assume an uncertainty of 0.1\% in the efficiency for $b$ tagging
and an uncertainty of 0.5\% in the uncertainty for charm tagging.
These values are based on an extrapolation of the results of $\ee\to
ZZ$ studies described in [46]. It would be valuable to confirm these
values with a full-simulation study at the higher statistics required
here, and that analysis is
 in progress. Note that, while these values affect our projections for the precision
 electroweak uncertainties given in Table \ref{tab:PEWresults}, they
 do not significantly affect the 
 uncertainties on Higgs boson couplings quoted in
 Table~\ref{tab:ILCHiggs}.

  For asymmetry
measurements, we must also discriminate $f$ from $\bar f$.  There 
is a correction due to sign flips, which must be estimated.  This
can be done using vertex, lepton, or kaon charges, collecting a 
sample of events with
non-contradictory charges ($-+$, $+-$).  To understand the effect of
sign flips, we can use also the sample of events with like charges
($--$, $++$).   This allows us to determine from the data themselves
the fraction of correctly reconstructed events and the fraction of
events that have suffered from migrations.  Thus, there is no need to
calculate a dilution factor from first principles and there is no
systematic error associated with dilution.  There is only a
statistical error that can be combined with other sources of
statistical error.

\section{Precision electroweak at 250 GeV from radiative return}  
\label{sec:return}

 In this report, we will discuss two
methods by which the ILC will improve on the precision measurements of
the $Z$ properties and couplings.   The highest precision measurements
will come from a GigaZ run, described in Sec.~\ref{sec:gigazaccel}.
We will present a detailed discussion of the GigaZ capabilities in
Sec.~\ref{sec:gigaz}.   

 However, we should not overlook the fact that
the ILC  running at 250~GeV will already produce a data set that will allow
substantial improvements of our knowledge of precision electroweak
observables.
One of the high-cross-section reactions at 250~GeV is the radiative
return to the $Z$,  $\ee\to Z\gamma$.  In this reaction, the $Z$ is
produced in the forward direction but still accessible to the ILC
detectors.  We will explain in a moment that the photon, which is
produced in 
the opposite forward
direction, does not need to be observed to provide a very clean event sample.
  The ILC program, with
2~ab$^{-1}$ of data, will produce roughly  77 million hadronic $Z$s
and 12 million leptonic $Z$s, a substantial increase over the event
sample of LEP.   Further, these events are produced with polarised
beams, so that, for polarisation observables, the event
sample to compare with is that of SLC.   The full power of the ILC
detectors can be used for flavor identification. 

We tag the signal events for the radiative return analysis 
 based on the polar angles 
of the two fermions from $Z\to f\bar{f}$. 
To describe the method simply, we will use the approximations that the fermions
 are massless and the photon is 
collinear to the beam directions. This is already quite close to
realistic, and the approximations can be relaxed with small
corrections.
Then let $E_i$ and $\theta_i$, $i=1,2$, 
 denote the energy and polar angle, respectively, of each final lepton
 or jet.   Transverse momentum conservation implies that 
$E_1\sin\theta_1=E_2\sin\theta_2$.
The fermion pair is boosted only in the beam direction.  
The boost
factor can be determined as
\beq
|\beta|=\frac{|E_1\cos\theta_1+E_2\cos\theta_2|}{E_1+E_2}
=\frac{|\sin(\theta_1+\theta_2)|}{\sin\theta_1+\sin\theta_2}.
\eeqn 
It is interesting that the $E_i$ cancel out, so 
 $\beta$ only depends on $\theta_1$ and $\theta_2$. 
The invariant mass of the fermion pair, $m_{12}$, can then be reconstructed as 
\beq
m^2 _{12}= \frac{1-|\beta|}{1+|\beta|} \cdot  s \ ,
\eeqn
where $\sqrt{s}$ is the center-of-mass energy. For the signal events
 we expect that $m_{12}$ peaks at $m_Z$ 
and, for $\sqrt{s} = 250$~GeV,  $|\beta|$ peaks at 0.76. The angles 
$\theta_1$ and $\theta_2$ can be measured very
precisely at the ILC detectors, so that the signal events can be
tagged 
without the need to observe the ISR photon.

This method was actually used at LEP2 ~\cite{ALEPH:1998ac}, 
though mainly for calibrating the beam energy due to the limited statistics. 
But at ILC250, we will expect $90$ million of such radiative events, a
factor of 5
 (100) more than the total number of $Z$ produced at LEP (SLC). 
 
\begin{figure}
\begin{center}
\includegraphics[width=0.70\hsize]{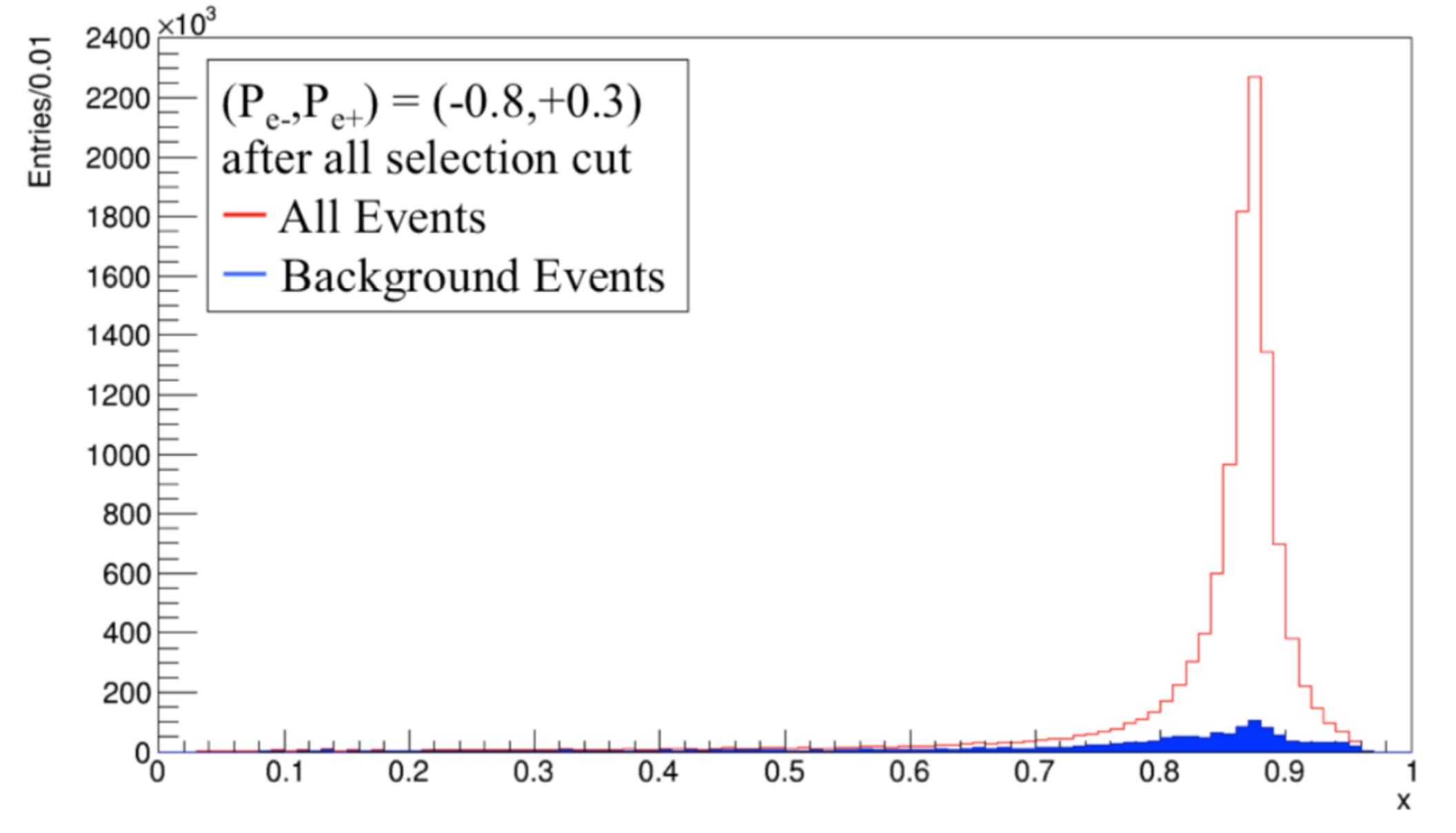}
\end{center}
\caption{Reconstructed distribution of $x\equiv\frac{2|\beta|}{1+|\beta|}$ 
for the signal $e^+e^-\to\gamma Z$, $Z\to q\bar{q}$ and from
background events 
that mimic this signal, 
at $\sqrt{s}=250$ GeV with an integrated luminosity of 250 fb$^{-1}$.}
\label{fig:beta_az}
\end{figure}

A fast simulation study has been performed for the $A_e$ measurement
using the 
$e^+e^-\to\gamma Z$, $Z\to q\bar{q}$ channels and the full SM 
background~\cite{Ueno:2019}. 
After all the selection cuts, the signal efficiency is 73\% and the
remaining
background events,  due to systems with
approximately the $Z$ mass from other processes,  are almost negligible,
as shown in Fig.~\ref{fig:beta_az}.  For the results shown, 
 realistic effects from finite fermion mass
and beam crossing angles have already been taken into account. 
The events in which the photon goes into the detector
have not been separated, but it should be straightforward to do,
 provided that they only contribute as a small fraction
of total events.
From the measured cross sections for the left- and right-handed beam
polarisations,
 $A_e$ can be determined from Eq.~\leqn{ALRmeas}.
The statistical error on $A_e$ for 2 ab$^{-1}$ data in the ILC250
scenario is 
estimated to be 0.00015. 
We can perform the same analysis using the $Z\to l^+l^-$ channels. 
The combined statistical error is expected
to be $\Delta A_e=0.00014$, a relative error of $\delta A_e=9.5\times
10^{-4}$.
This is a factor of 10 improvement over the current uncertainty  on $A_e$.
Many systematic errors in the cross section measurement cancel out in
the measurement of this asymmetry. 
The dominant systematic error for $A_e$ will  come from the
uncertainty in  $P_{eff}$.  In Sec.~\ref{sec:beampol}, we have
explained that, through the measurement of processes with large polarisation
asymmetries  such as $e^+e^-\to WW$, the relative systematic error on
$A_e$ can be reduced to $3\times 10^{-4}$.

In principle, the value of $A_e$  also depends on 
 the CM energy in the $\ee\to Z\gamma$ reaction.  The polarisation
asymmetry actually measured in this reaction has the form~\cite{Rowson:2001cd}
\beq
A_{obs} =A_e+\Delta A,
\eeqn
where $\Delta A$ is a correction due to interference between the
contributions  to the  $\ee\to
f\bar f \gamma$ from the resonant diagram with an intermediate  $Z$ and the
nonresonant diagram with an intermediate  $\gamma$. At the $Z$ pole,
the interference term has significant energy-dependence, requiring
excellent knowledge of the CM energy.   This will be an issue in
Sec.~\ref{sec:weakmixing}.    However, for the radiative return
process at 250~GeV, the dependence $\Delta A_e/\Delta E_{CM}$ is 3
orders of magnitude smaller, allowing us to safely ignore the
systematic
error from the beam energy uncertainty.

For $A_f$ measurements other than $A_e$, we need to measure the
left-right forward-backward asymmetry defined in Eq.~\ref{Afmeas}. A
dedicated simulation study for $A_f$ ($f=b/c/\mu/\tau$) has not yet been performed. Nevertheless we can estimate the signal
efficiency in two steps based on existing simulation analyses.
 The first step is to tag the signal events as from radiative return,
just as in the $A_e$ measurement. The second step is to identify
the
 flavor and charge of the fermion. 
For example,  the efficiency for  the $A_b$ measurement can be estimated
to 
be $73\%\times 40\%$ in which the 73\%, for 
tagging the hadronic radiative return event, is from fast simulation
analysis described above~\cite{Ueno:2019}, 
and the 40\%, for $b$-tagging and $b$ charge identification, is from a
full simulation
 analysis described in ~\cite{Roman:2019a}. 
The statistical error of $A_b$ is then estimated to be $\Delta
A_b=0.00053$, a relative uncertainty of $\delta A_b=5.7\times 10^{-4}$.
Similarly, the efficiencies for $A_c$, $A_\tau$ and $A_\mu$ can be
derived from full simulation 
results in~\cite{Roman:2019b, Daniel:2019, Deguchi:2019tvp}.  These
are estimated to be 
 $73\%\times 10\%$, 80\%,  and 88\%,  respectively. 
Their statistical errors are summarized in Table~\ref{tab:PEWresults}.
The dominant systematic error is expected to come from the uncertainty
in the  effective beam polarisation $P_{eff}$. 

The measurements of 
$R_f (1/R_f)\equiv \Gamma(Z\to f\bar{f})/\Gamma(Z\to$~hadrons),
for $f=b/c$ ($f=e/\mu/\tau$),
are  simpler to describe, since we only need to measure the total rate
for each flavor 
without the
need to identify the charge. The signal efficiencies can be 
estimated based on the same analyses cited above
\cite{Ueno:2019, Roman:2019a, Roman:2019b, Daniel:2019, Deguchi:2019tvp}. 
The expected statistical errors are summarized in
Table~\ref{tab:PEWresults}. The 
dominant systematic errors would
come from the uncertainties in the 
 flavor-tagging efficiencies,  estimated in Sec.~\ref{sec:PEW} to be 0.1\% for $f=e/\mu/\tau/b$ and 
0.5\% for $f=c$.  

Noting that  $R_e$ is expected to be improved by a factor of 2 over
the current uncertainty ~\cite{Tanabashi:2018oca}, it is interesting
to convert this to a value of the quantity
$\Gamma_e\equiv\Gamma(Z\to e^+e^-)$, which is a useful input to the
 SMEFT global fit for Higgs boson couplings that will be
discussed in Sec.~\ref{sec:SMEFT}.  $\Gamma_e$ can be derived from 
the measurements of the cross section of $Z$ to hadrons
$\sigma_{\mathrm{had}}$,  the $Z$ total width $\Gamma_Z$,  and $R_e$, with
the uncertainty estimated as 
\beq
\delta\Gamma_e=\frac{1}{2}\delta\sigma_\mathrm{had}
\oplus\frac{1}{2}\delta\Gamma_Z\oplus\frac{1}{2}\delta R_e \ ,
\eeqn
where $\delta$ denotes a relative uncertainty: $\delta A = \Delta
  A/A$. 
With the current uncertainties for $\sigma_\mathrm{had}$ and
$\Gamma_Z$ 
from ~\cite{Tanabashi:2018oca}, 
and expected uncertainty for $R_e$ at ILC250, we expect the precision
for $\Gamma_e$ to decrease to 
$\delta\Gamma_e=0.86\times 10^{-3}$.  This improves over the current
relative 
uncertainty  on $\Gamma(Z\to \ell^+\ell^-)$ of $1.02\times 10^{-3}$ and
also allows us to relax the assumption of lepton universality in this
input to the SMEFT fit.

\section{Precision electroweak from the GigaZ program}
\label{sec:gigaz}

\newcommand{\sintw}{\sin^2 {\rm \theta^_{eff}}}
\newcommand{\sintwb}{\sin^2 {\rm \theta^{b}_{eff.}}}

A further improment of precision electroweak observables is possible in the 
GigaZ program described in Sec.~\ref{sec:gigaz}.   As shown in Table~\ref{tab:pollumiNZ}, the GigaZ program will produce about $5\times10^9$ $Z$
events.  This is  equivalent to about 250 times more than has been collected at LEP by all four experiments, thus promising improvements of electroweak observables by more than one order of magnitude.  
The machine would be operated with polarised beams with a degree of polarisation of $|80\%|$ for electrons and $|30\%|$ for positrons.

\subsection{Measurements of the weak mixing angle}
\label{sec:weakmixing}

We first discuss measurements of the effective weak mixing angle 
$\sin^2\theta_{eff}$,
defined in Eq.~\ref{sstweffdef}, and, more general, the leptonic left-right 
asymmetries $A_\ell$ for $\ell = e,\mu,\tau$.  At a linear collider with polarised beams the effective weak mixing angle  can be extracted in several ways but notably by measuring the left-right asymmetry $A_{LR}$, Eq.~\leqn{ALRmeas}.  Using this method,  SLD achieved the highest-precision single measurements of $\sstw$, even though LEP had collected about 30 times more luminosity.

At GigaZ, using all hadronic decay modes of the $Z$, the statistical error on $A_{LR}$ will be a few times $10^{-5}$.  The measurement will then be dominated by the 
systematic error on the polarisation.  As we have explained in Sec.~\ref{sec:beampol},
we expect a systematic error on the beam polarisation of 0.05\% for a realistic assumption of
 0.25\% of the  precision of the polarimeters. Positron polarisation plays a crucial role in reaching this low level of uncertainty. 

A precise measurement of $A_{LR}$ at the $Z$ pole requires also excellent control over the value  of the beam energy. The observed polarisation asymmetry has 
a strong energy-dependence due to the interference of the $s$-channel $Z$ and $\gamma$ diagrams:
 $dA_{LR}/dE_{CM}\approx 2\times 10^{-5}/{\mathrm {MeV}}$. But we   have argued in 
Sec.~\ref{sec:beamenergy} that the beam energy in GigaZ can be measured, by a combination of methods, to a precision of 1~MeV. This is of similar size to the 
statistical error.  We note that this $A_{LR}$ measurement  is specifically a measurement of 
$A_e$. 

The values of $A_\mu$ and $A_\tau$ can also be improved at GigaZ by measuring 
the corresponding left-right forward-backward asymmetries, Eq.~\leqn{ALRmeas}. 
Note that, for a lepton species, the left-right forward-backward asymmetry at the $Z$ is 7  times larger than that unpolarised forward-backward asymmetry and less subject to radiative corrections.   It is interesting to test lepton universality by comparing these
quantities  to 
the precisely measured value of $A_e$.   The systematic error due to  the 
polarisation cancels out in the ratios, so $A_\mu$ and
$A_\tau$ can be compared to $A_e$ with a relative uncertainty of about
0.02\%. 
 
 The higher
statistics  available from GigaZ 
will of course improve the
measurements of $R_{\ell}$  for each lepton species.   The systematic
errors are small.  Also, these are  due to knowledge of the acceptance, so
it can be assumed that these errors scale with luminosity. To obtain
the estimates in Table~\ref{tab:PEWresults},  we have simply rescaled the LEP results
given in \cite{ALEPH:2005ab}.

The absolute precision on $\sin^2\theta_{eff}$ of  $1.3\cdot10^{-5}$ expected from GigaZ is nearly one order of magnitude better than the precision of the present world average $\sstw$~\cite{Tanabashi:2018oca} and only a factor three worse than that  claimed  for FCCee~\cite{Abada:2019zxq}. This  is reminiscent of the LEP/SLC scenario.  It is worth recalling some details of the measurement of $A_e$ at circular colliders.  The best method is to use a wonderful formula from LEP:  the $\tau$ polarisation at the $Z$ varies with the $\tau$ production angle $\theta$ according to~\cite{Eberhard:1989ve}
\beq
P_\tau(\cos\theta) = - {A_\tau (1 + \cos^2\theta) + 2 A_e \cos\theta \over 
 (1 + \cos^2\theta) + {8\over 3} A^\tau_{FB} \cos\theta } \approx 
        A_\tau + { 2\cos\theta\over (1 + \cos^2\theta)} A_e \ . 
\eeq{LEPmagic}
Since $A_e$ controls the $\cos\theta$ asymmetry in this formula, 
 it is in practice somewhat better 
determined than $A_\tau$.  This  gives the best determination of $\sin^2\theta_{eff}$.  The dominant systematic error in this technique is the 
uncertainty in the conversion of the measured energies of $\tau$ decay products to 
 the underlying $\tau$ polarisation.  This is complicated by the fact that all $\tau$ decay modes receive feed-down from other modes for which the observed energy spectrum of the 
visible decay products has a different dependence on the $\tau$ polarisation.
In the LEP era, this cross-contamination was about 10\% in each mode, but the modelling of $\tau$ decays was  understood 
 well enough to constrain this error on $A_\tau$, $A_e$ to be  less than  1\% (relative error).    For FCCee, this understanding must be improved by two orders of magnitude.  Some difficulties in achieving this are explained  in \cite{Vincter}.

\subsection{Measurements of heavy quark production}
\label{sec:gigaZheavy}

Other important observables of the $Z$ pole experiments are the $Z$ couplings to the heavy quarks $b$ and $c$.  In this section, we discuss the measurement of these 
quantities and some physics implications of those measurements.   This subject 
is treated more comprehensively in  \cite{Irles:2019xny}. Note, though, that \cite{Irles:2019xny} supposes an unpolarised positron beam. 

We first present estimates of the precision of the determinations of $R_b$ and $R_c$ and of $A_b$ and $A_c$.   The basic methods for these measures were described in 
outline in Sec.~\ref{sec:PEW}.  For the $b$ observables, the efficiencies that determine the statistical errors are derived from the study of $\ee\to b\bar b$ presented in 
\cite{Bilokin:2017lco}.   For $c$, the statistical errors are extrapolations of those 
presented in \cite{ALEPH:2005ab}.  

The systematic errors bring in some more subtle points.
Thanks to the excellent vertex detector and the small beam size,  the ILC 
experiments are much closer to the SLD detector than the LEP detectors, and so one 
might take the SLD heavy quark analyses as a starting point. 
The relevant references for this are \cite{Abe:2005nqa} and \cite{Abe:2004hx}. One finds  that, apart from Monte Carlo statistics, there is
not a single  dominant source of systematic error. Instead, the total systematic  error is composed of a number of small contributions. It is safe to assume that most of these contributions will be controlled to a sufficient level at the time of GigaZ, either by improved understanding of QCD or by higher-statistics measurements of 
$\ee\to q\bar q$ processes.   As an example, one large source of systematic error for $c$ quark observables is the
uncertainty from  gluon splitting to a $c \bar{c}$ pair. Consulting the OPAL analysis in \cite{Abbiendi:1999sxa},  one finds that the uncertainty of the splitting fraction was limited by statistics that did not allow for a sufficient reduction of the background from $b$-quark pairs. At  GigaZ, it will be possible to take advantage of the much higher
statistics in $q\bar q$ production both at GigaZ and at ILC250, and also  the detector will be superior to the OPAL detector. It is therefore justified to assume that the gluon splitting can be controlled to a much better level than it was possible for OPAL. Since the measurements not statistics-limited, we can study the influence of other QCD effects by comparing to a fiducial region in which the heavy quark jets are approximately back-to-back.


Following these considerations, the dominant error source for $A_{b}$ is given by the uncertainty of beam polarisation. In case of $A_{c}$ we assume that the error of sources other than beam polarisation will roughly equal the error of beam polarisation. In case of $R_b$, the general improvement of the measurements justifies an improvement of the systematic error by a factor of five. (This 
improvement was already
found in the studies for the TESLA Technical Design Report~\cite{AguilarSaavedra:2001rg}). For $R_c$, it is justified to assume that the component of the systematic error that does not improve with statistics will be improved from SLC by a factor of about two. 

We have already pointed out in Sec.~\ref{sec:PEW} that the method for determining 
heavy quark forward-backward asymmetries will be much improved from that of LEP using the large sample of double-tagged events.   The systematic error from this source in the LEP experiments will become a statistical error that is continuously improvable. 

\begin{figure}
\centering
\includegraphics[width=0.7\textwidth]{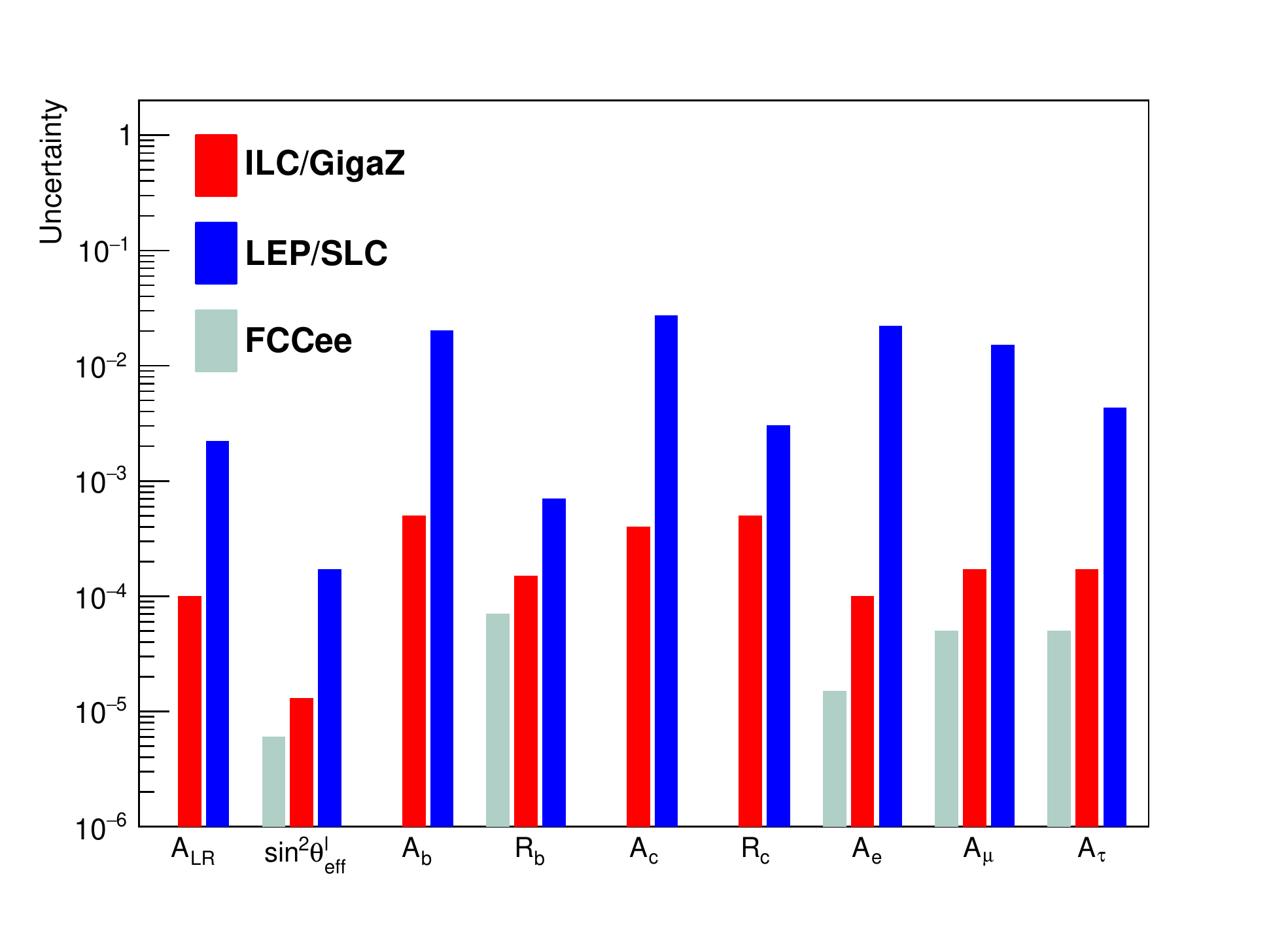}
\caption{\label{fig:alr-gigaz} \sl Summary of the precision achievable at GigaZ compared  with LEP/SLC results~\cite{ALEPH:2005ab} and FCCee projections~\cite{Abada:2019zxq} for observables and derived quantities that are described in the text.}
\end{figure}

Figure~\ref{fig:alr-gigaz} summarises the precisions expected at  GigaZ for  the heavy quark observables.    These results are also presented in Table~\ref{tab:PEWresults} in Appendix A.

There are two important physics motivations for studying the heavy quark 
couplings to the $Z$ beyond the general idea of finding higher-precision tests of the SM.   The first is that the largest deviation of the precision electroweak 
observables from the  SM predictions observed in the LEP/SLC program 
 involves the $b$ system. Assuming, following the SM expectation, that $A_b$ is close to 1, 
one can extract $A_e$ from a measurement of
the $b$ forward-backward asymmetry, using Eq.~\leqn{AFBmeas}.  At LEP, this 
determination gave results that differ from the arguably more direct measurements of $A_e$ from the left-right asymmetry and the $\tau$ polarisation asymmetry by about 3.5 standard deviations.  Actually, there is a lack of rapport that involves the 
three quantities $A_{FB}^b$, $R_b$, and $A_e$ that frustrates theoretical 
explanations~\cite{Takeuchi:1994zh}.  This issue calls for a new set of experimental 
measurements. 

\begin{figure}
\centering
\includegraphics[width=0.7\textwidth]{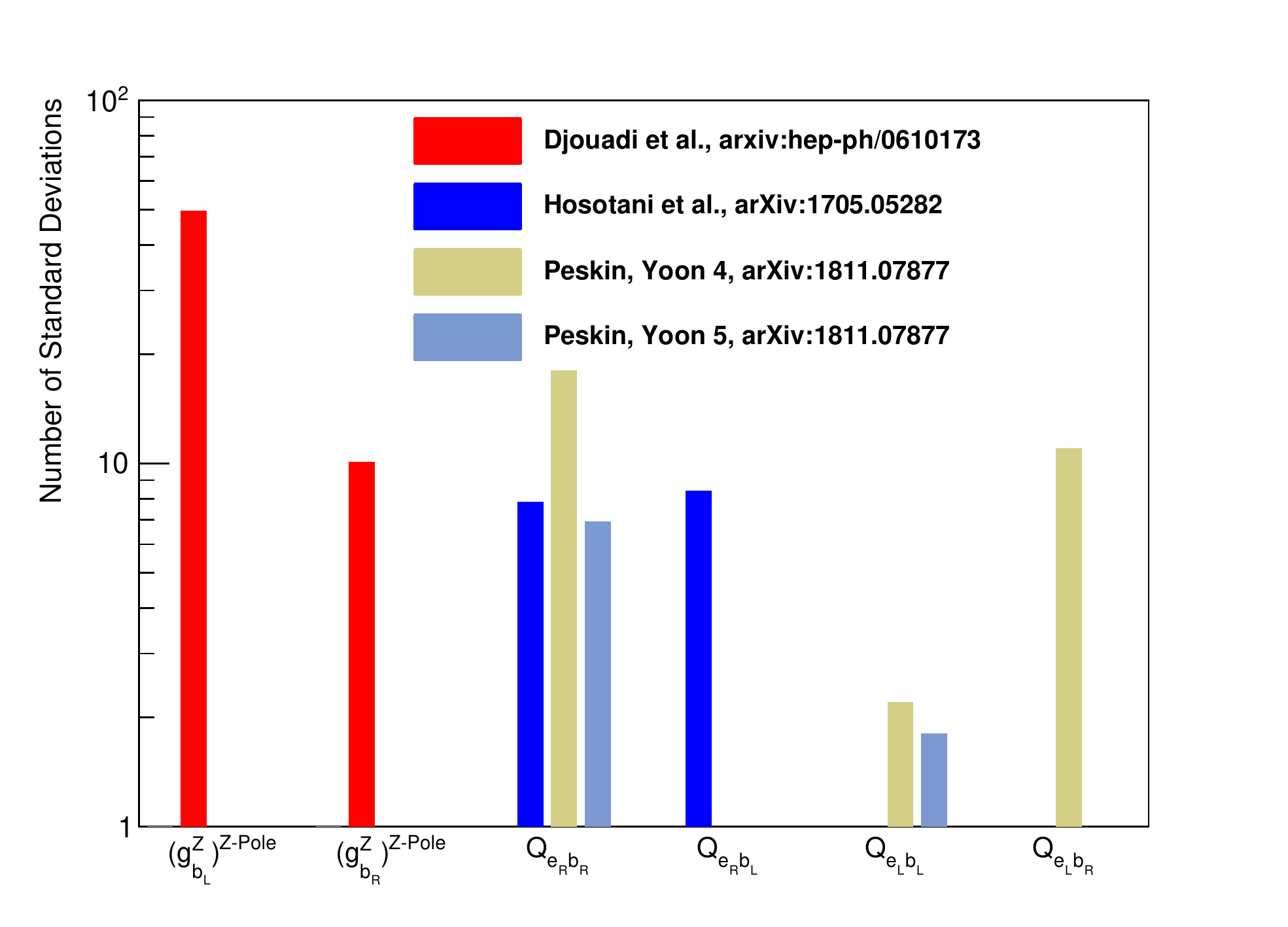}
\caption{Visibility of deviations from the SM predictions in $g_{b_j}^Z$ and the helicity amplitudes $Q_{e_i b_j}$, in  standard deviations,  from combined ILC250/GigaZ running, expected from new physis models with Randall-Sundrum extra dimensions~\cite{ Djouadi:2006rk,Funatsu:2017nfm,Yoon:2018xud}.}
\label{fig:gigaz-rs} 
\end{figure}

The second is the real possibility that the $Z$ couplings to the $b$ quark are 
 altered by new physics.  The $b_L$ is in the same electroweak multiplet as the $t_L$,
so if the  top quark acquires its large mass from strong dynamics in the Higgs sector, 
the $b_L$ also must feel 
the effects of this new strong sector.   The direct coupling of the $b$ to the Higgs sector can be made small since $m_b \ll m_t$, and the couplings of the $b$ to 
photons and gluons are restricted by Ward identities, so the one place where such 
corrections are allowed to show up is in the $b$ coupling
 to weak-interaction bosons.
The $b$ quark can also couple preferentially to $Z'$ bosons associated with the 
Higgs strong interactions.  All of these features are explicitly realized in Randall-Sundrum extra-dimensional 
models of the Higgs sector~\cite{Djouadi:2006rk,Funatsu:2017nfm,Yoon:2018xud}.

These expectations can be tested through measurements of $\ee\to b\bar b$.  For some models, the effects are already large enough to see at ILC250.  Using 
polarisation and the forward-backward asymmetry, the ILC250 can measure the 
four helicity amplitudes associated with $b$ couplings to $Z$ and $Z'$,
\beq
   Q_{e_i b_j} = Q^\gamma_e Q^\gamma_f + {g^Z_{e_i} g^Z_{b_j}\over s - m_Z^2} 
   + {g^{Z'}_{e_i} g^{Z'}_{b_j}\over s - m_{Z'}^2} \ . 
\eeq{bhelicity}
for $i, j = L, R$.   The second term can include effects of $Z$-$Z'$ mixing~\cite{Djouadi:2006rk}.    For example, at ILC250, the quantity 
  $Q_{e_L b_R}$, which has a SM value of about 0.45, can be measured with a
precision of  $\Delta Q_{e_L b_R} = 5 \times 10^{-4}$.   If there is a deviation from the 
SM, we will want to resolve whether it comes from the $Z$ couplings or the couplings 
to higher resonances.  That could be done with a second measurement at the $Z$ pole.
To match the ILC250 determination, the $Z$ pole measurment should reach a precision 
of 0.5\%, achievable at GigaZ but a factor of 10 beyond the current precision from LEP. 
The sensitivity of models to combined ILC250/GigaZ running is shown in 
Fig.~\ref{fig:gigaz-rs}.
The model of \cite{Funatsu:2017nfm} predicts similar perturbation of the helicity 
amplitudes for the light fermions, so it is interesting to pursue these measurements also
for the lighter flavors.

\subsection{Measurements of total and partial widths}

Using the improved knowledge of the beam energy discussed in Sec.~\ref{sec:beamenergy}, it will be possible to improve the systematic error on the 
width of the $Z$  to about 1~MeV, with negligible statistical error~\cite{AguilarSaavedra:2001rg}.  This would be an 
improvement over the LEP uncertainty by more than a factor of 2.    Given this improvement
and the GigaZ improvement in $R_e$, the relative uncertainty in the quantity $\Gamma(Z\to \ee)$ highlighted at the end of Sec.~\ref{sec:return} improves to 
$0.56\times 10^{-3}$.

In all, we see that the GigaZ program is surprisingly powerful.   It has the capability 
to improve all of the $R_f$ snd $A_f$ precision observables by a factor of 10 from 
their current LEP and SLC values.  In some cases, we obtain a much larger improvement.
The program strongly benefits from the use of polarised beams
 and the high level of control that these 
give us over the limiting systematic errors.  The full set of projected uncertainties 
for the GigaZ program is given in Table~\ref{tab:PEWresults}.

\section{4-fermion processes}
\label{sec:fourfermion}

In addition to precision tests of the SM in $Z$ boson couplings, the
ILC will bring new tests of the SM in four-fermion interactions, which
will be measured with precision at 250~GeV and higher energies.
Within the SM, fermion pair production cross sections are very well
understood and computed to part per mille accuracy. Precision
measurements of these processes at $\ee$ colliders can recognize small
deviations from these predictions.  In this way, it is possible to
test both for the presence of new $s$-channel electroweak resonances
and for four-fermion contact interactions that represent the
low-energy effective description of new electroweak sectors. 

Several features of $\ee$ collisions make this type of search
especially powerful.  First, one knows that the initial state is
$\ee$, and it is possible to distinguish flavors in the final state.
Also, in the approximation of ignoring initial- and final-state
masses, the differential cross sections for 100\% polarised
$e^-_Le^+_R$ and $e^+_Le^-_R$ beams take the
form
\beqa
{d \sigma\over d \cos \theta} (e^-_Le^+_R \to f {\bar f}) &=&
   \Sigma_{LL}(s)  \ (1 + \cos\theta)^2 + \Sigma_{LR}(s)
   (1-\cos\theta)^2 \CR
{d \sigma\over d \cos \theta} (e^-_Re^+_L\to f {\bar f}) &=&
   \Sigma_{RL}(s)  \ (1 - \cos\theta)^2 + \Sigma_{RR}(s)
   (1+\cos\theta)^2
\eeqa{basiccs}
where $\Sigma_{LL}$, $\Sigma_{RL}$ refer to $f_L\bar f_R$ production
and $\Sigma_{LR}$, $\Sigma_{RR}$ refer to $f_R\bar f_L$ production.
This means that, with polarised beams,  each process gives 4
independently measurable coefficients that can provide tests of the
SM. 

This section will discuss ``universal'' parameters of
  four-fermion interactions and parametrizations appropriate to
  production of light flavors.   Pair-production of $b$ and $t$ quarks
  within the SMEFT brings in a large number of operator coefficients;
  fits to this larger set of parameters are reviewed in
  Sec.~\ref{sec:bt}. 

\subsection{Searches for  $Z'$ bosons}

We first discuss the search for new $s$-channel $Z^\prime$ resonances.
In Table~\ref{tab:Zprime}, we present exclusion and discovery 
limits for various types of $Z^\prime$ bosons 
that are considered in the literature. A commonly used 
metric is the reach for the Sequential Standard Model (SSM) $Z^\prime$ 
whose couplings are assumed to be  identical to the couplings of the $Z$ boson of the SM. 
Another benchmark is  the ALR model,  which features a boson that couples to the right-handed $SU(2)$ 
acting on SM fermions with the same strength as the weak-interaction left-handed $SU(2)$. 
An actual $Z^\prime$ would have couplings orthogonal to the couplings
of the Z, so actually, both the SSM and the ALR models are straw
men. With this in mind, we also quote results for $Z'$ bosons 
 found in $E_6$ grand unified theories 
that extend the SM, canonically taken as the linear combinations
$\psi$,
 $\chi$ and $\eta$ of two bosons from the center of $E_6$ 
orthogonal to the SM directions.
The limits in the table are based on an analysis
 of $e^+e^−\to f\bar{f}$, $f= e/\mu/\tau/b/c$, at 250 GeV 
using the ILD detector model and the full
 simulation framework described in Sec. 6 and 7 of~\cite{Bambade:2019fyw}, 
assuming a data set of 2~ab$^{-1}$. This analysis is described in 
some detail in~\cite{Deguchi:2019tvp}; 
we also include information from the studies in~\cite{Roman:2019a,Roman:2019b,Daniel:2019}. 
The background events in all of those channels are essentially
negligible.
 The signal efficiencies in $e$, $\mu$ and $\tau$ 
channels are rather high, respectively 97\%, 98\% and 90\%. For $b$
and $c$ channels, 
mainly due to charge identification,
the efficiencies are much lower, 29\% and 7\%, respectively. The
exclusion and 
discovery limits for $Z^\prime$ are obtained
based on a $\chi^2$ fit to the measured differential cross sections 
$\mathrm{d}\sigma/\mathrm{d}\cos\theta$ where $\theta$
is the polar angle of the fermion. Systematic errors are taken into
account 
in the fit; they are assumed to 
0.1\%, 0.1\%, 0.2\%, 0.2\% and 0.5\% respectively for $e/\mu/\tau/b/c$ channels.
We have extrapolated these results to ILC500 with 4 ab$^{-1}$ and to
ILC1000 with 8~ab$^{-1}$.
The results for higher CM energies go beyond the 
current reach of the LHC and eventually surpass the reach of the HL-LHC. 
It is important to note that, in the event of a discovery of a
$Z^\prime$ at the HL-LHC, 
the ILC will provide complementary information 
that will be essential in pinning down the nature of the new
resonance.

\begin{table}[t]
\begin{center}
\begin{tabular}{l|cc|cc|cc}
                    & 250 GeV,  &  2 ab$^{-1}$    &  500 GeV,  &   4 ab$^{-1}$   &   1 TeV,   &  8 ab$^{-1}$   \\
    Model       &  excl.        &   disc.              & excl.          &    disc.            &  excl.        &  disc.   \\ \hline                
    SSM         &  7.8          &    4.9                &  13            &   8.4                &  22           &   14      \\
    ALR          &  9.5          &    6.0                &  17            &   11                 &  25           &   18      \\
    $\chi$       &  7.0          &    4.5                &  12            &   7.8                &  21           &   13      \\
    $\psi$       &  3.7          &    2.4                &  6.4           &   4.1                &  11           &   6.8      \\
    $\eta$       &  4.2          &    2.7                &  7.3           &   4.6               &  12           &   7.9     
\end{tabular}
\end{center}
\caption{Projected limits on $Z^\prime$ bosons in standard scenarios, 
from the full simulation study of $e^+e^−\to f\bar{f}$ described in the text. 
The values presented, given in TeV, are the 95\% exclusion limits and 
the 5$\sigma$ discovery limits for the successive stages of the ILC program up to 1 TeV.}
\label{tab:Zprime}  
\end{table}

\subsection{Measurement of ``universal'' four-fermion interactions}

\begin{figure}
\begin{center}
\includegraphics[width=0.70\hsize]{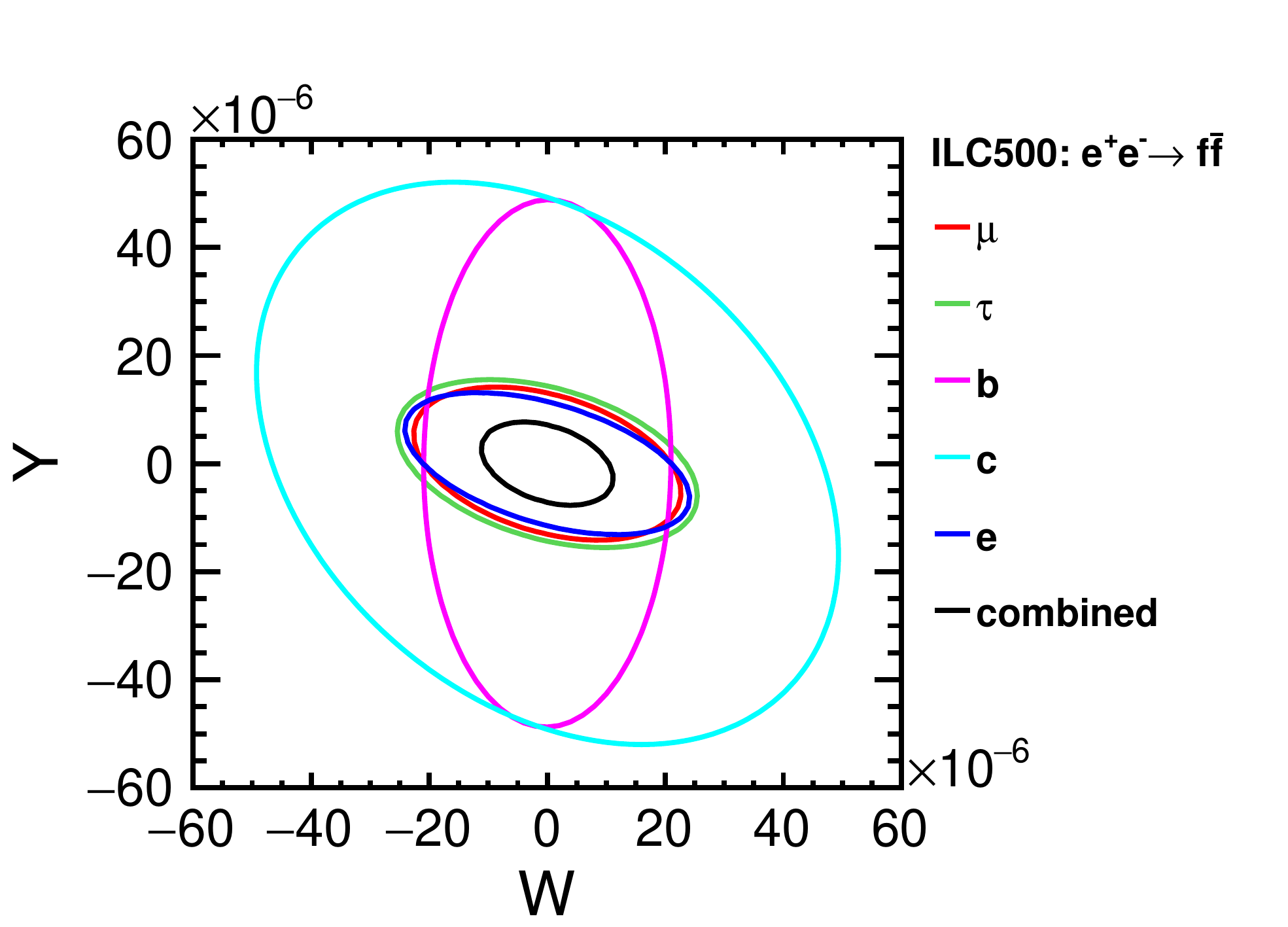}
\end{center}
\caption{68\% confidence contours in the ${\bf W}, {\bf Y}$  plane for ILC at 500 GeV, 
from the $e^+e^-\to f\bar{f}$ analysis described in the text. 
The colored contours show the contributions from each flavor to the final combined limit.}
\label{fig:WandY}
\end{figure}

The same analyses for probing the $Z^\prime$ can be recast into a set of measurements of the 
``universal'' four-fermion interactions characterized by the parameters
 ${\bf W}$ and ${\bf Y}$ defined in~\cite{Barbieri:2004qk, Farina:2016rws},
\beq
   \L = \L_{SM} - {g^2  {\bf W }\over 2 m_W^2} J_{L\mu}^a J^{a\mu}_L - {g^{\prime
     2} {\bf Y}\over 2 m_W^2} J_{Y\mu} J^{\mu}_Y \ , 
\eeq{LWYdef}
where $g$ and $g'$ are the SM coupling constants for $SU(2)$ and
$U(1)$ and  $J_{L\mu}^a$, $ J_{Y\mu}$ are the corresponding gauge
currents. 
The combined results are shown in Table~\ref{tab:WandY} for the three energy stages of the ILC. 
It is also interesting to see how the results for each 
flavor contribute to the final constraints. This is shown in Fig.~\ref{fig:WandY}. 
It is important to note that the beam polarisation plays a central role in
this analysis to disentangle the effects from ${\bf W}$ and ${\bf Y}$. 
In the last line of Table~\ref{tab:WandY}, we show the comparable results for 4 ab$^{-1}$ of data at 500 GeV with no beam polarisation. 
Not only are the results poorer, but also the correlation between
${\bf W}$ and ${\bf Y}$ is significantly increased.
\begin{table}[t]
\begin{center}
\begin{tabular}{lccc}
   $\sqrt{s}$     &   $\Delta {\bf W}$      &   $\Delta {\bf Y}$  &
                                                                    $\rho$   \\ \hline 
   HL-LHC  &       $15\times 10^{-5}$  &   $20\times 10^{-5}$  &
                                                                 -0.97 \\ \hline
    ILC250     &     $3.4\times 10^{-5}$  & $2.4\times 10^{-5}$  & -0.34 \\
    ILC500     &     $1.1\times 10^{-5}$  & $0.78\times 10^{-5}$  & -0.35 \\
    ILC1000   &     $0.39\times 10^{-5}$  & $0.27\times 10^{-5}$  & -0.38 \\        
   \hline
    500 GeV, no beam pol.     &     $2.0\times 10^{-5}$  & $1.2\times
                                                           10^{-5}$ 
 & -0.78 \\    
\end{tabular}
\end{center}
\caption{Projections for 1-$\sigma$ errors on ${\bf W}$ and ${\bf Y}$
 from a 2-parameter fit to data 
on $e^+e^−\to f\bar{f}$ from the analysis described in the text. 
The assumed luminosities are those described in
Section~\ref{sec:accel}. The projection for HL-LHC (3~ab$^{-1}$) is 
 based on the neutral current analysis described in 
\cite{Farina:2016rws}, in particular, Fig.~2 of that paper. } 
\label{tab:WandY}  
\end{table}

\subsection{Measurement of general four-fermion interactions}

A more specific description of the constraints on the 4-fermion contact interactions is given by the ``compositeness parameters'' 
as defined in~\cite{Tanabashi:2018oca}, 
\beq
 \L = \L_{SM}  \pm \L_{LL} \pm \L_{LR} \pm \L_{RL} \pm \L_{RR} 
\eeq{fullL}
with 
\beqa
\L_{LL}&=&\frac{g^2_\mathrm{contact}}{2\Lambda^2}\sum_{j}\eta_{LL}^{j}
(\bar e_L\gamma_\mu e_L)(\bar{\psi}^j_L\gamma^\mu\psi^j_L), \CR
\L_{LR}&=&\frac{g^2_\mathrm{contact}}{2\Lambda^2}\sum_{j}\eta_{LR}^j
(\bar e_L\gamma_\mu e_L)(\bar{\psi}^j_R\gamma^\mu\psi^j_R), \CR
\L_{RL}&=&\frac{g^2_\mathrm{contact}}{2\Lambda^2}\sum_{j}\eta_{RL}^j
(\bar e_R\gamma_\mu e_R)(\bar{\psi}^j_L\gamma^\mu\psi^j_L), \CR
\L_{RR}&=&\frac{g^2_\mathrm{contact}}{2\Lambda^2}\sum_{j}\eta_{RR}^j
(\bar e_R\gamma_\mu e_R)\bar{\psi}^j_R\gamma^\mu\psi^j_R), 
\eeqa{eqn:composite}
where $j$ indexes the final-state fermion flavor. At the ILC,
individual  $eeff$-type contact interactions can be measured for each fermion species
$f=e/\mu/\tau/b/c$.  In addition, the four parameters $\eta_{LL}$,
$\eta_{RR}$, $\eta_{LR}$ and $\eta_{RL}$ 
can in principle be determined simultaneously using 
the differential cross section measurements with  polarised beams. We
have not yet performed a fit for all $\eta$ parameters simultaneously.
Instead, following the conventional approaches
in~\cite{Tanabashi:2018oca},  we give projections for the 95\% exclusion limits on scale $\Lambda$ 
for several cases of assumed $\eta$ values,
\beqa
\Lambda&=&\Lambda_{LL}^{\pm} ~\mathrm{for}~ (\eta_{LL},\eta_{RR},\eta_{LR},\eta_{RL})=(\pm1,0,0,0), \CR
\Lambda&=&\Lambda_{RR}^{\pm}~ \mathrm{for} ~(\eta_{LL},\eta_{RR},\eta_{LR},\eta_{RL})=(0,\pm1,0,0), \CR
\Lambda&=&\Lambda_{VV}^{\pm}~ \mathrm{for} ~(\eta_{LL},\eta_{RR},\eta_{LR},\eta_{RL})=(\pm1,\pm1,\pm1,\pm1), \CR
\Lambda&=&\Lambda_{AA}^{\pm}~ \mathrm{for} ~(\eta_{LL},\eta_{RR},\eta_{LR},\eta_{RL})=(\pm1,\pm1,\mp1,\mp1),
\eeqa{eqn:comsim}
and $g^2_\mathrm{contact}/(4\pi)=1$. These results are presented in
Table~\ref{tab:composite}.  The first group of limits assumes that the contact interactions 
are universal for all fermion species, the following groups give the
results for each specific final-state fermion species. 
The comparable limits from LEP were about 8 TeV, that is, about 40 times the CM energy. 
With the increased luminosity of the ILC and the use of polarisation,
we expect to be 
sensitive to $\Lambda$ values of over 200 times the CM energy.
\begin{table}[p]
\begin{center}
\begin{tabular}{lcccc}
   $\sqrt{s}$     &   $\Lambda_{LL}$      &   $\Lambda_{RR}$  & $\Lambda_{VV}$   &  $\Lambda_{AA}$    \\ \hline \hline
   universal $\Lambda$'s & & & & \\ \hline
    ILC250     &     108  &  106  &  161  &  139 \\
    ILC500     &     189  &  185  &  280  &  240 \\    
    ILC1000   &     323  &  314  &  478  &  403 \\ \hline\hline
   $e^+e^-\to e^+e^-$ & & & & \\ \hline
    ILC250     &     71  &  70  &  118  &  71 \\
    ILC500     &     114  &  132  &  214  &  135 \\    
    ILC1000   &     236  &  232  &  376  &  231 \\ \hline\hline   
   $e^+e^-\to \mu^+\mu^-$ & & & & \\ \hline
    ILC250     &     80  &  79  &  117  &  104 \\
    ILC500     &     134  &  133  &  198  &  177 \\    
    ILC1000   &     224  &  222  &  332  &  296 \\ \hline\hline   
   $e^+e^-\to \tau^+\tau^-$ & & & & \\ \hline
    ILC250     &     72  &  72  &  109  &  97 \\
    ILC500     &     127  &  126  &  190  &  168 \\    
    ILC1000   &     215  &  214  &  321  &  286 \\ \hline\hline   
   $e^+e^-\to b\bar{b}$ & & & & \\ \hline
    ILC250     &     78  &  73  &  103  &  106 \\
    ILC500     &     134  &  124  &  175  &  178 \\    
    ILC1000   &     226  &  205  &  292  &  296 \\ \hline\hline   
   $e^+e^-\to c\bar{c}$ & & & & \\ \hline
    ILC250     &     51  &  52  &  75  &  68 \\
    ILC500     &     90  &  90  &  130  &  117 \\    
    ILC1000   &     153  & 151  & 220  &  199 \\ \hline\hline   
\end{tabular}
\end{center}
\caption{Projected 95\% CL limits, in TeV,  on the compositeness scales
  defined in~\cite{Tanabashi:2018oca},
from $e^+e^-\to f\bar{f}$ analysis described in the text. 
In all cases, 
the limits from constructive ($\Lambda^+$) and destructive
($\Lambda^-$)
 interference are identical. 
The first group of numbers assumes that the $\Lambda$ parameters are independent of flavor. 
The succeeding groups show the limits for the reactions with specific final state flavors.}
\label{tab:composite}  
\end{table}

\section{ Pair production of $b$ and $t$ quarks}
\label{sec:bt}

We have already pointed out in Sec.~\ref{sec:gigaZheavy} 
 that electroweak couplings of the top and bottom quark are of special interest
for a number of reasons.  In that section, we pointed to possible improvements
in the $Z$ pole coupling to $b\bar b$.  The story of these couplings
is actually more general, but that general analysis requires
measurements at higher energies.    As was pointed out in that
section, new physics can influence heavy quark pair production both
through modification of the $Z $ and $\gamma $ couplings and through
the addition of four-fermion interactions mediated by new heavy gauge
bosons or other particles of a strongly-coupled Higgs sector. The
latter effects were discussed for $b$ quarks in
Sec.~\ref{sec:fourfermion}.  It is
possible to 
discuss the full  variety of these effects in a common framework by
making use of Standard Model Effective Field Theory (SMEFT).
Recently, the SMEFT analysis of heavy quark electroweak couplings and
the constraints from future $\ee$ experiments have been analyzed in 
\cite{Durieux:2018ekg,Durieux:2019rbz}.  In this section, we will
review some results of that work.

\subsection{Measurement of the top quark mass}

The top quark mass is one of the key parameters of the Standard Model
and must be determined experimentally. A precise determination
requires exquisite control over experimental and theoretical
effects. In this section, we give a brief review of the  measurement
of $m_t$ at ILC, with references 
to the relevant literature.

The current world average, with contributions from the Tevatron and
LHC experiments is $m_{t} =$ 172.9 $\pm$
0.4~GeV~\cite{Tanabashi:2018oca}. The experimental uncertainties are
expected to improve to approximately 200~MeV at the HL-LHC. 
Significant theoretical  work is still required to connect this quoted  value of
$m_t$  to a well-defined short-distance top quark mass
at that level of   precision~\cite{Azzi:2019yne}.

At $\ee$ colliders, a very precise measurement of the top
quark mass, with a total uncertainty of approximately 50~MeV, is
possible by scanning the centre-of-mass energy through the $t\bar{t}$
production
threshold~\cite{Simon:2019axh,Abramowicz:2018rjq,Vos:2016til}. The
dominant uncertainty is expected to be the theoretical
uncertainty~\cite{Simon:2016pwp}.  The theoretical expression 
for the threshold shape is now known to N$^3$LO order, but this still 
leaves a small residual theoretical uncertainty~\cite{Beneke:2015kwa}. A measurement with a
precision that surpasses that of the HL-LHC legacy measurement is also
possible from the ILC running at $\sqrt{s}=$ 500~GeV, by
taking advantage of radiative 
$e^+e^- \rightarrow t\bar{t}\gamma$ events~\cite{Abramowicz:2018rjq}.

\subsection{Measurement of top quark  electroweak couplings}


The third-generation quarks play a special role in many extensions of the
Standard Model. Several proposed extensions predict large deviations of the
electroweak couplings of the bottom and top quark from the Standard Model
value. A precise characterization of the $e^+e^-\rightarrow b\bar{b}$ and
$e^+e^-\rightarrow t\bar{t}$ processes at an electron-positron collider
can probe such models to very high scales.
These measurements are particularly powerful in composite-Higgs models
and Randall-Sundrum models with additional (warped) space-time
dimensions~\cite{Agashe:2006at,Richard:2014upa}, with the discovery
potential extending up to scale of tens of TeV~\cite{Durieux:2018ekg}.


The $e^+e^- \rightarrow b\bar{b}$ process was studied extensively at
LEP and SLC and the electroweak precision tests at the $Z$ pole remain
the most powerful constraint on the $Zb\bar{b}$ vertex today. Operation
of the ILC at $\sqrt{s}=$ 250 GeV allows to determine
the $Z$-boson and photon vector and axial couplings to b-quarks.
With an integrated luminosity of 2 $ab^{-1}$ the form factors in the
general Lagrangian can be measured at the few-per-mille
precision~\cite{Irles:2019int,Bilokin:2017lco}. This implies an order of
magnitude improvement with respect to the LEP combination in the determination
of the right-handed coupling of the $b$-quark. 
The results at $\sqrt{s}=$250 GeV complement the $Z$-pole data from
the LEP/SLC experiments and GigaZ~\cite{Irles:2019xny}.


The top quark escaped scrutiny at the previous generation of electron-positron$\ee$
colliders. Measurements at the Tevatron and the LHC have characterized many
of its properties. Rare associated production processes, such as
$pp \rightarrow t\bar{t}X$ with $X$ a $Z$-boson, photon or
Higgs boson, yield direct access to the neutral-current electroweak
interactions and the top quark Yukawa coupling, while single top production
and top decay probe the $tWb$ vertex. A fit of the top quark
effective field theory to the LHC data has recently been
performed~\cite{Hartland:2019bjb}. The constraints on the operator coefficients
that affect the top quark electroweak couplings are still rather weak.
A combined fit of the bottom and top quark sector to LHC and LEP/SLC data
yields slightly improved limits~\cite{Durieux:2019rbz}.

\begin{figure}[t]
    \centering
    {\includegraphics[width=0.98\textwidth]{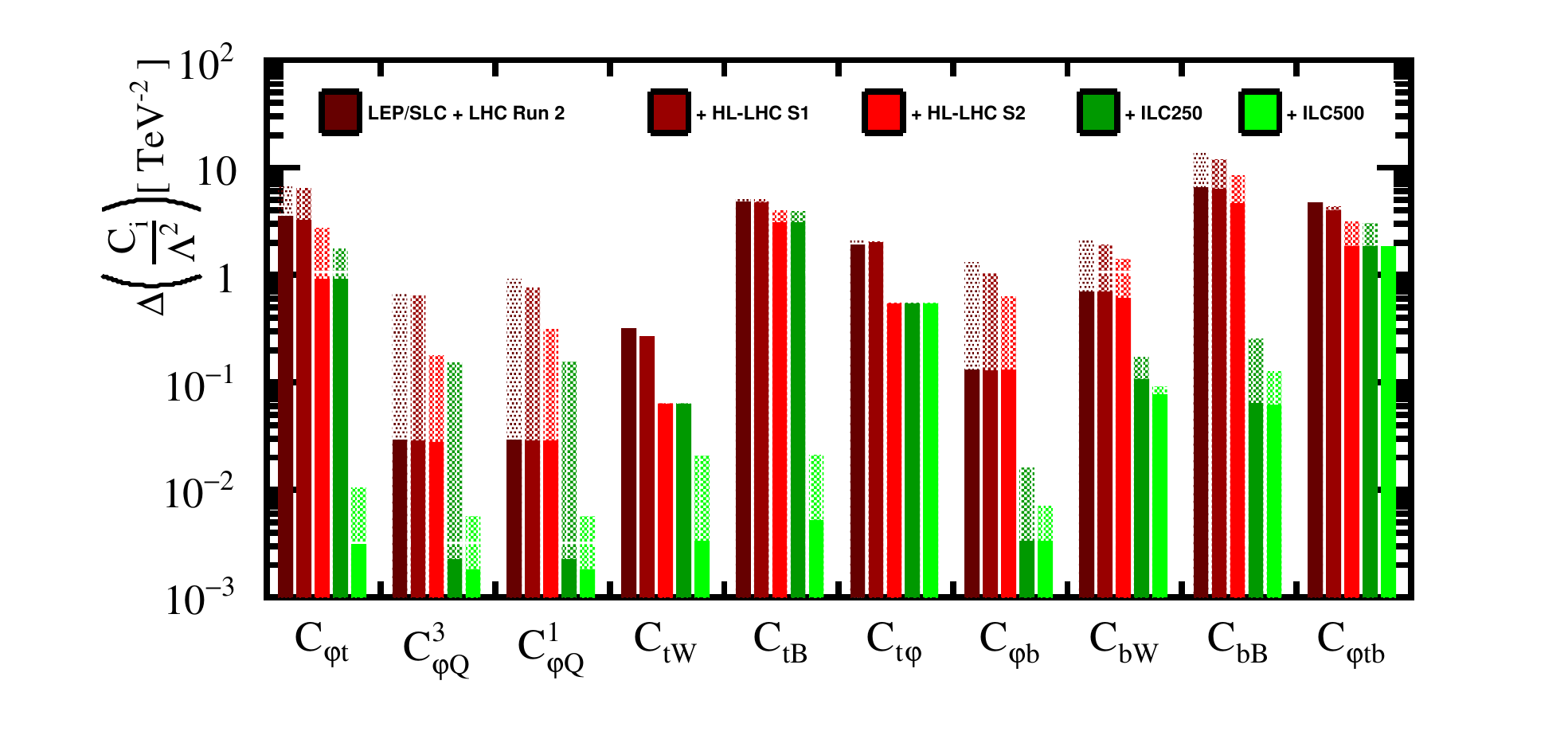}}
    \caption{Prospects for the precision of the Wilson coefficients in
      future high-luminosity operation of the LHC and at a high-energy
      $e^+e^-$ collider, from~\cite{Durieux:2019rbz}.
      The  figure shows the results of a fit to 10 coefficients of SMEFT
      operators that modify the electroweak 
couplings of the $b$ and $t$ quark. 
The solid section of the bars represents the individual constraints,
where each parameter is fitted in isolation; the full length indicates
the marginalized constraint in a ten-parameter fit.}  
    \label{fig:manhattan_plot}
\end{figure}


The $e^+e^- \rightarrow t\bar{t}$ process opens up at CM energies greater
than twice the top quark mass. Measurements of the cross section and
forward-backward asymmetry for two configurations of the beam polarisation
determine the left- and right-handed couplings
with sub-\% precision~\cite{Amjad:2015mma}. Measuring several further
observables allows us to overconstrain a global fit of all operator
coefficients
that directly affect these couplings~\cite{Durieux:2018tev}.

The prospects for the SMEFT fit of the top and bottom quark electroweak couplings
are summarized in Fig.~\ref{fig:manhattan_plot}. This plot shows the result of a fit to a set of
10 operator coefficients corresponding to shifts in the $b$ and $t$ electroweak vertices.  The current bounds are based
on LEP/SLC data and LHC results~\cite{Durieux:2019rbz}.    The HL-LHC results are presented for two scenarios, S1 and S2.  In the S2
scenario,   LHC results are extrapolated to the complete HL-LHC luminosity
assuming that experimental systematic uncertainties scale with the inverse
of the integrated luminosity and that theory uncertainties improve by a
factor two. The ILC prospects are based on the full simulation studies
\cite{Irles:2019int} and \cite{Amjad:2015mma,Abramowicz:2018rjq}.  The effect of four-fermion operators involving $b$ and $t$
is not included in this fit, though those should be considered as part of a full SMEFT analysis.    This case is discussed in the following 
section. 


\subsection{Measurement of the top quark Yukawa coupling}


The top quark Yukawa coupling links top and Higgs physics.   Several
measurements of Higgs boson production and decay rates, in particular
$gg \rightarrow H$ production and $H \rightarrow gg$,
$H\rightarrow \gamma \gamma$ and $H\rightarrow Z \gamma$ decay, are sensitive
to this coupling. Under a number of model-dependent assumptions,
the Yukawa coupling can be extracted with good precision from the
Higgs fit (see Section~\ref{sec:SMEFT}
and, for instance, Ref~\cite{Boselli:2018zxr}).

The most robust measurement is obtained in associated production of a top quark
pair and a Higgs boson. An ILC run with 4~ab$^{-1}$ at 550 GeV is
expected to reach 3.3\% precision, similar to the precision
envisaged in the HL-LHC S2 scenario~\cite{Cepeda:2019klc}. Operation of the ILC at a
centre-of-mass energy of 1 TeV improves the precision of the Yukawa
coupling by a further factor two~\cite{Durieux:2019rbz}, to 1.6\%.
These results, both for ILC and LHC, are obtained from a 1-parameter fit to the $t\bar t h$
cross section, assuming that the other couplings contributing to this
cross section have their SM values.

In the context of the SMEFT, many dimension 6 operators specifically involving b and t affect the 
$t\bar t h$ production cross section.  At hadron colliders, there are about 30 distinct operators 
that must be considered; at $\ee$ colliders, there are 17.  
A robust determination of the operator coefficient  $C_{t\varphi}$ that shifts the
Yukawa coupling requires precise constraints on the coefficients of these other operators. 
A recent global fit of the
top-quark sector on LHC data~\cite{Hartland:2019bjb} using the complete set of operators finds that the
marginalized fit result on $C_{t\varphi}$ is significantly poorer than
the individual limit, since  the operators that affect
the QCD interactions of the top quark (in particular the $q\bar{q}t\bar{t}$ operators) are still only weakly constrained.
 At 500~GeV, the ILC data set still cannot 
 determine the full set of operator coefficients needed in the $\ee$ analysis, because of degeneracy between the effects of 
four-fermion operators and shifts of the electroweak couplings.   This degeneracy is broken by data from higher energies.  
The ILC including data at 1~TeV  can provide a robust
result in a global fit, because at that point the data with different beam polarisations and
at two CM energies over-constrain the full set of relevant SMEFT parameters~\cite{Durieux:2019rbz}.



\subsection{Requirements for $b$ and $t$ quark measurements}

The study of the bottom and top quarks leads to several specific
 requirements on the
accelerator and operating scenario.
Beam polarisation is a key tool to disentangle the photon
and $Z$-boson contributions to the $b\bar{b}$ and $t\bar{t}$ pair production
processes~\cite{Amjad:2015mma}. The study~\cite{Durieux:2018tev} finds a 15\%
degradation of the overall EFT constraints without positron polarisation
and a 50\% degradation if no beam polarisation is available at all.
Polarisation plays an even larger role in the analysis of the 
$e^+e^-\rightarrow b\bar{b}$ process, since final-state polarisation is
accessible only through measurement of the angular distribution with
polarised beams. 

While the $e^+e^- \rightarrow b\bar{b}$ is accessible at
the first energy stage of the ILC at 250 GeV, top quark pair production
requires a CM  energy of at least 350 GeV. Measurements well
  above threshold are desireable to escape uncertainties from threshold
  effects and to increase the sensitivity to axial-vector top quark couplings.
The threshold  for associated production of
a top quark pair with a Higgs boson lies at approximately 500 GeV.
Operation at still higher energy is very effective in  constraining
operators whose 
effects that grow with CM energy~\cite{Durieux:2018tev}.
A robust and global characterization of all operators in the effective
field theory requires an extended programme with operation at
250 GeV, $\sqrt{s}=$ 500-550 GeV, and 1 TeV.


Third-generation quarks pose stringent requirements on the detector design
and selection and reconstruction algorithm. Efficient and clean identification
of jets from the hadronization of a bottom quark ($b$-tagging) is crucial
for these analyses. The measurement of the
forward-backward asymmetry in the $e^+e^- \rightarrow b\bar{b}$ analysis
and in the fully hadronic final state in $e^+e^- \rightarrow t\bar{t}$
production moreover requires to distinguish jets from the fragmentation
of $b$ and $\bar{b}$ quarks. In the analysis of~\cite{Bilokin:2017lco}
this is achieved through a combination of
vertex charge measurements and identification of kaons with low and
  medium momentum.
The requirements of an effective $b/c$ and $b/\bar{b}$ separation are
important drivers in the design of the vertex detector. The analysis of
$b\bar{b}$ and $t\bar{t}$ pair production is therefore an important
benchmark for the design of the experiments.

The main challenge in the reconstruction of $t\bar{t}$ and $t\bar{t}H$
events is the large jet multiplicity. With up to eight jets in the
final state,
jet clustering becomes a major challenge~\cite{Boronat:2016tgd}. For
CM energies of 1 TeV and beyond the boost of the top quark
is such that dedicated reconstruction algorithms are
required~\cite{Abramowicz:2018rjq}.

\section{SM EFT Higgs coupling fit at ILC}
\label{sec:SMEFT}

At the ILC, the most powerful method for determining the Higgs
  boson couplings uses a global fit of SMEFT parameters  to $\ee$
  observables.
The default fit includes 16 dimension-6 operators, 
4 SM constants and 2 parameters for Higgs to 
invisible and other exotic decays. By making use of the input measurements
from electroweak precision observables, 
$e^+e^-\to WW$, Higgs observables at the HL-LHC and Higgs observables 
at the ILC, it is possible to fit all the 
parameters simultaneously. The details of the SMEFT formalism and the input
measurements are explained
 in~\cite{Bambade:2019fyw, Barklow:2017suo, Barklow:2017awn}. 
From the expected HL-LHC results, we use only the ratios of branching
ratios for the Higgs boson decays to $\gamma\gamma$, $ZZ^*$,
$Z\gamma$, and $\mu^+\mu^-$;  these ratios have a clear
model-independent
interpretation.
Please note that the ILC beam polarisations play a very important role. 
At 250 GeV, the capabilities of 2 ab$^{-1}$ data with beam 
polarisations are almost as same as that of 
5 ab$^{-1}$ of data without beam polarisation. 
Here we only give the updated information with respect to that 
presented in~\cite{Bambade:2019fyw}.    

First of all, we extend the SMEFT fit to include the measurements at 
1 TeV. The estimated statistical errors for the input
Higgs measurements~\cite{Asner:2013psa} and $e^+e^-\to WW$ 
measurements~\cite{Rosca:2016hcq} at 1 TeV 
based on full detector simulation studies are given
in Table~\ref{tab:higgserrors} and~\ref{tab:WWerrors}. 
Second, we include the improved electroweak measurements
from radiative return events at 250 GeV, discussed in Sec.~\ref{sec:return},
into our default fit. Previously, in~\cite{Bambade:2019fyw}, we only
included the factor of 10 improvement in 
$A_\ell$ in the S2$^*$/S2 scenarios. 

The improvement of precision electroweak measurements, even at the
level of the radiative return analysis, allows us to take another
important step.   The projections
in \cite{Bambade:2019fyw} and \cite{Barklow:2017suo} made use of the
assumption of lepton universality for their input values of $A_\ell$
and $\Gamma_\ell$.  Actually, though, the Higgs boson
cross sections depend only on $A_e$ and $\Gamma_e$.   We have
explained at the end of Sec.~\ref{sec:return} that the radiative return
analysis allows us to improve $A_e$ to a level that is 10 times better
than that of the current $A_\ell$ and to improve $\Gamma_e$ to a level 
15\% better than that of  the current $\Gamma_\ell$.   This allows us to
remove the assumption that the $Z$ couplings to the three lepton
species are identical.    This assumption can be separately tested in the
radiative return analysis and at GigaZ, as we have explained in
Secs.~\ref{sec:return} and \ref{sec:gigaz}, but it is now decoupled
from the fit for Higgs boson couplings.

\begin{figure*}[t]
\begin{center}
\includegraphics[width=0.8\hsize]{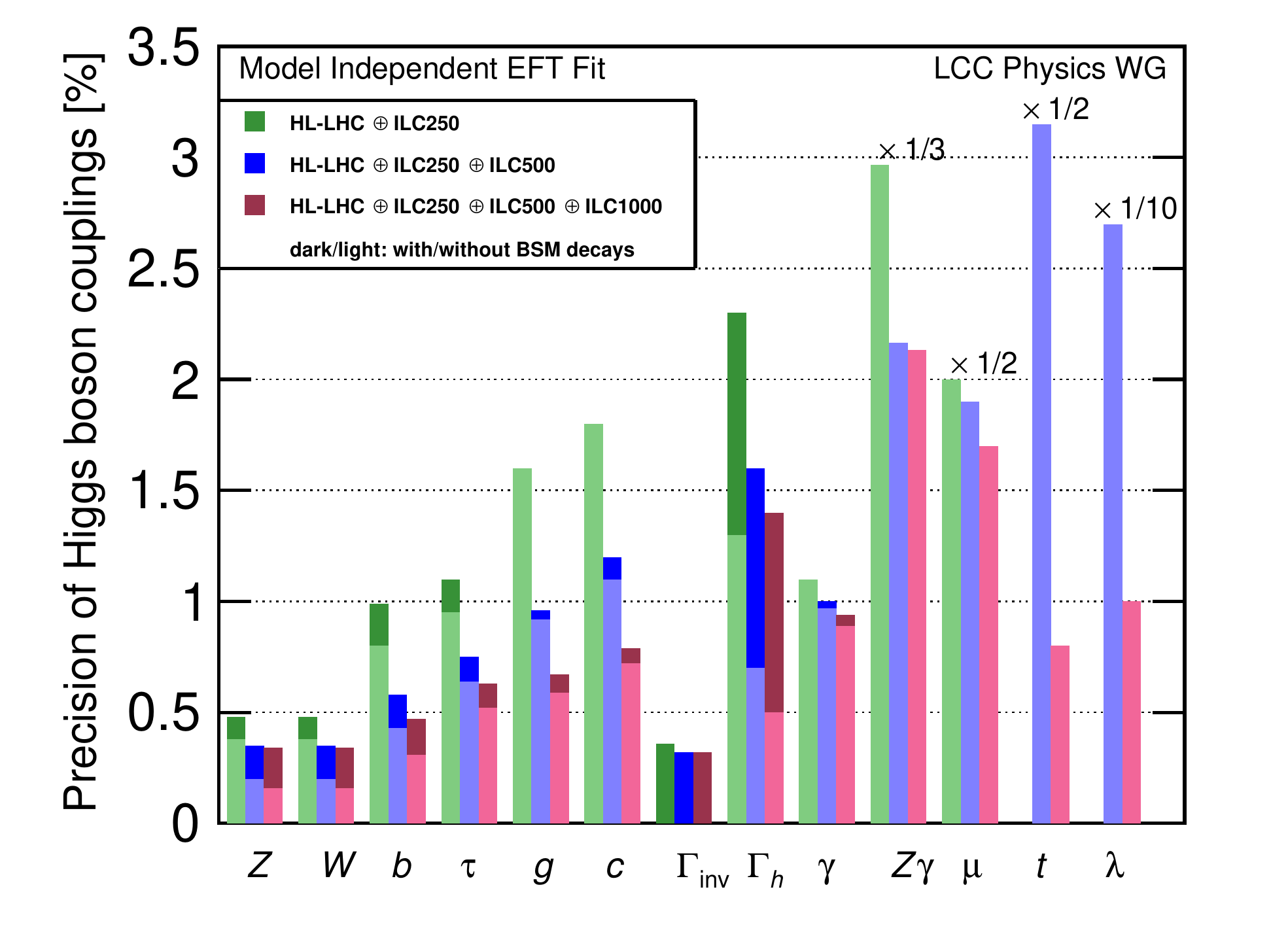}
\caption{Projected Higgs boson coupling uncertainties for ILC250,
  ILC500, and ILC1000, also  incorporating  results expected from the
  HL-LHC, based on the SMEFT analysis described in the text.   The
  darker bars show the results allowing invisible and exotic Higgs
  decay channels; the lighter bars assume that these BSM decays are
  not present.  The column $\lambda$ refers to the $HHH$ coupling.  In the last four columns, 
  all bars are rescaled by the indicated factor.}
\label{fig:Higgsresults}
\end{center}
\end{figure*}
%

Table~\ref{tab:ILCHiggs} in Appendix A gives the updated uncertainties in the Higgs
couplings
 for the ILC at 250 GeV and 500 GeV, and the new results
at 1 TeV. Systematic errors are taken into account, as discussed
in~\cite{Bambade:2019fyw}. 
The global SMEFT fit discussed by the ECFA Higgs
@ Future Colliders working group~\cite{deBlas:2019rxi} made the
assumption that there are no 
invisible or other exotic decays of the Higgs boson.  To facilitate a
comparison with that group, we also give the 
projected uncertainties with this  assumption in addition to the
projections from our default fit. Please note that the effect is quite
significant.  In particular, the very small errors
projected for the $W$, $Z$, and $b$ couplings projected for ILC and
other facilities depend on the assumption of no exotic Higgs decays.
The projections in the table are shown graphically in Fig.~\ref{fig:Higgsresults}.

Dedicated $Z$ pole running of the ILC with GigaZ
would improve the uncertainties  $A_e$ and $\Gamma_e$ by a factor of
20 and 2,  respectively,  over current uncertainties, as we have
discussed
in Sec.~\ref{sec:gigaz}.
It is interesting to see how much improvement this brings  in the
Higgs boson couplings determined by the SMEFT fit.  In
Table~\ref{tab:ILCHiggsGigaZ2}
we show  the effect of this improvement over the projections quoted in  
Table~\ref{tab:ILCHiggs}.  It turns out that the effect is minor and
disappears almost entirely for the ILC when 500~GeV data is included.
A similar effect for other facilities was noted in
\cite{Bambade:2019fyw}.  To make the differences more visible, we have
computed the projections in Table~\ref{tab:ILCHiggsGigaZ2} using the
assumption that the Higgs boson has no exotic decays. 

\begin{table}[t]
\begin{center}
\begin{tabular}{lccc}
   coupling     &   2~ab$^{-1}$\ at 250      &   + 4~ab$^{-1}$ at 500  &
                                                                +8~ab$^{-1}$
  at 1000 \\ \hline 
$hZZ$            &             0.35 / 0.38 &            0.20 /   0.20    &
                 0.16  /  0.16   \\ 
$hWW$            &          0.35 / 0.38  &            0.20  /  0.20      &                                                
                              0.16 /  0.16 \\ 
 $hbb$            &     0.79  / 0.80  & 0.43 /0.43   &  0.31 / 0.31 \\ 
$h\tau\tau$    &          0.94 / 0.95 &   0.63 / 0.64 & 0.52 / 0.52 \\ 
$hgg$                  &  1.6 /  1.6    &   0.92  / 0.92    
 & 0.59 /  0.59 \\ 
$hcc$         &   1.7  /  1.8&  1.1   /   1.1 &    0.72 /   0.72  \\ 
$h\gamma\gamma$ &  1.0  / 1.1 &   0.95  /  0.97 & 0.88  / 0.89 \\ 
$h\gamma Z$     &  8.5 /  8.9 &   6.4 /  6.5 &  6.3  / 6.4 \\
$h\mu\mu$ &  4.0  / 4.0 &  3.8  /  3.8 &     3.4  /  3.4     \\ 
$htt$  &   ---     &      6.3   &      1.6      \\ 
$hhh$  &  ---    &   27 &      10    \\ \hline 
$\Gamma_{tot}$ & 1.3 / 1.3 & 0.70 / 0.70 &  0.50  /  0.50\\  
\end{tabular}
\end{center}
\caption{  Projected uncertainties in the Higgs
  boson couplings for the ILC250, ILC500, and ILC1000, with 
  precision LHC input.   All values
  are given in percent (\%).   The two values in each field are 
  for fits with and without Giga-$Z$ running.  Both values are computed
  under the assumption of no invisible or untagged Higgs boson  decays.}
\label{tab:ILCHiggsGigaZ2}  
\end{table}

It is worth noting that in the global SMEFT fit discussed here we 
include all of the the operators that are relevant to the Higgs couplings at
the leading order. 
High-precision measurements of deviations of the Higgs couplings from
the SM
 will need to
consider also next-to-leading order effects.   At this order
more operators are expected to
contribute, in particular the top quark operators that are discussed
in Sec.~\ref{sec:bt}. The Higgs and top measurements that will be
available at the
 HL-LHC and ILC will become 
more intimately connected when we discuss
 the global SMEFT fit at NLO. Please be invited to read the paper~\cite{Jung:2019}.

\section{ Higgs self-coupling}
\label{sec:self}

At $\ee$ colliders, the higgs self-coupling can be measured directly
in double Higgs production in the reactions $\ee\to Zhh$ and $\ee\to
\nu\bar\nu hh$.    The first of these reactions can be studied at the
ILC at 500~GeV; the second requires higher energy to obtain a
sufficient
event sample.    The $Zhh$ process was analysed
in \cite{Takubo:2009ws,Tian:2010np,Duerig:2016dvi}
through full-simulation studies of this process in the decay 
modes $hh\to (b\bar
b)(b\bar b)$ and  $hh\to 
(b\bar b)(W+W^-)$.   For a 4~ab$^{-1}$ data set, these results 
extrapolate to a precision of 
27\%~\cite{Fujii:2015jha}, if the self-coupling has a value close to
the SM expectation. 
The extraction of the self-coupling  was based on a
1-parameter fit varying the Higgs self-coupling only.  But also,  it was shown in
\cite{Barklow:2017awn}  that this uncertainty is essentially unchanged
in the context of a SMEFT analysis including all possible dimension-6
operators, since the additional relevant operator coefficients  are
strongly constrained
by single-$h$ measurements at the ILC.

The same framework has been used to perform  full-simulation studies of the cross
section measurement for $\ee\to
\nu\bar\nu hh$ at 1~TeV, again using the $hh\to (b\bar
b)(b\bar b)$ and  $hh\to 
(b\bar b)(W^+W^-)$ decay modes~\cite{Tian:2013yda}.
   Extrapolating these results
to a data set of 8~ab$^{-1}$, we find an estimated precision of 10\%
on the Higgs self-coupling~\cite{Bambade:2019fyw}, again for the case
in which the self-coupling is close to the SM value.
   The authors of \cite{Barklow:2017awn}
believe that this uncertainty estimate will  also turn out to be highly
model-independent, though the analysis has not yet been completed.

\begin{figure}[tb]
  \begin{center}
   \includegraphics[width=0.7\hsize]{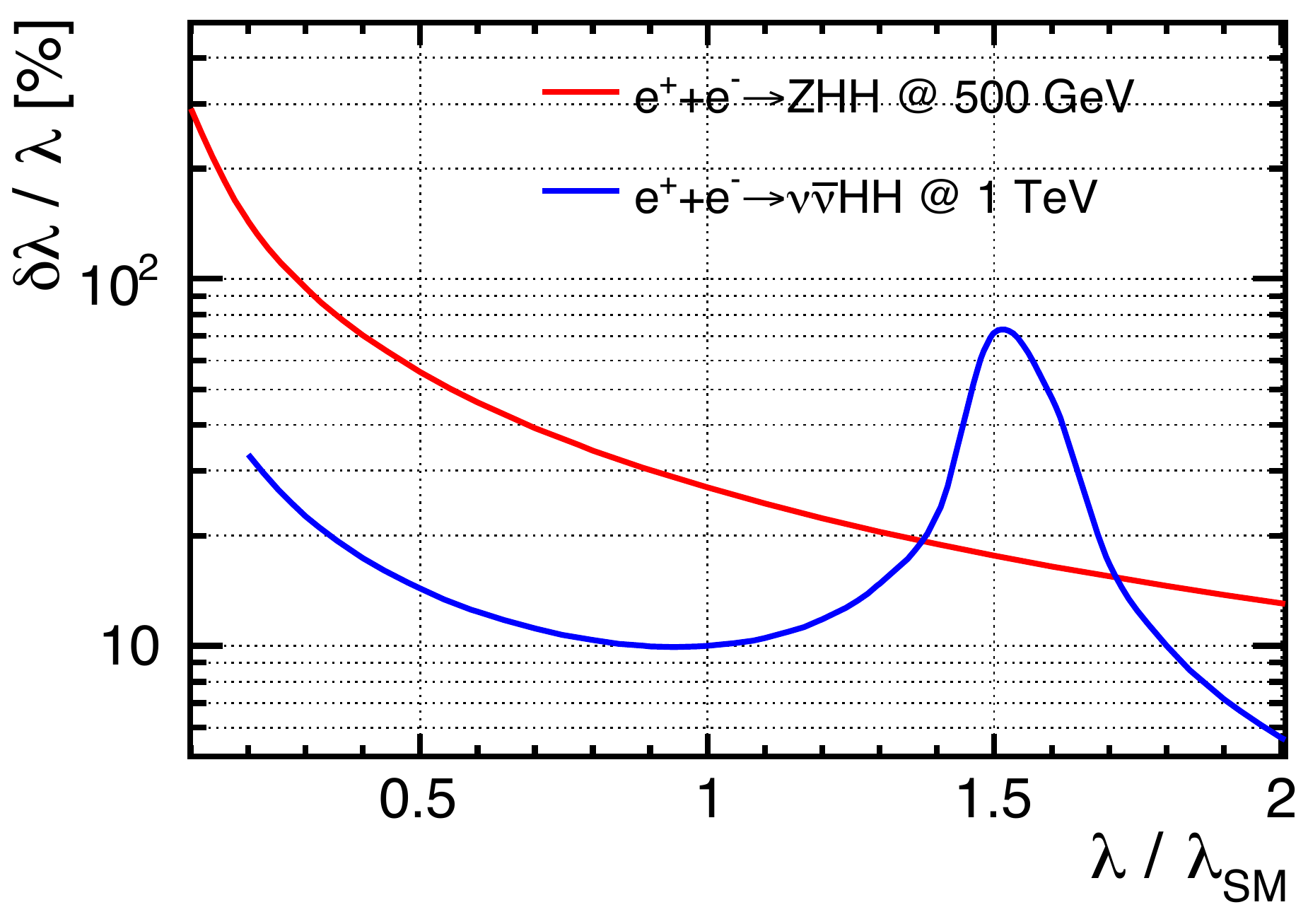} 
\end{center}
  \caption{The expected precisions of the Higgs self-coupling
    $\lambda$ at the ILC as a function of the assumed value of
    $\lambda$,  for the
    reactions $e^+e^-\to Zhh$ ~(red) and
for $e^+e^-\to\nu\bar{\nu}hh$ ~(blue).  The luminosities assumed are: 
 4~ab$^{-1}$ at 500~GeV and  4~ab$^{-1}$ with 
 $(P_{e^-},P_{e^+}) = (-80\%, +20\%)$ at 1~TeV. }
  \label{fig:HHHSensitivity}
\end{figure}

In theoretical models, especially in models of electroweak
baryogenesis, the Higgs self-coupling can be enhanced over its SM
value by a factor of 2.   It is therefore important to take account of
the fact that the uncertainty on the Higgs self-coupling varies with
the assumed value of this coupling.   In all processes, the digram
containing the 
self-coupling appears in interference with other SM diagrams.  
The uncertainty  on the Higgs self-coupling decreases in the case of
constructive interference but increases in the case of destructive
interference.  (The latter situation is well known for $gg\to
hh$ at hadron colliders.)   An advantage of $\ee$ collider studies is
that the two reactions $Zhh$ and $W$ fusion reactions have opposite
signs for the interference term.    This effect is shown in
Fig.~\ref{fig:HHHSensitivity},
which shows that, whether the self-coupling increases or decreases
from its SM value, the combination of the two $\ee$ reactions shows
robust performance, reaching precisions below 15\%  in the combination
for any values of the self-coupling larger than 0.4 times the SM value.

It should be emphasized, both for $t\bar t h$ production and for both
modes of $hh$ production, that the numbers we have quoted here are based on
our current reconstruction algorithms applied to full-simulation data.
In all three cases, the efficiencies of these algorithms leave
substantial room for improvement.  We thus expect that further study
and especially experience will real data will improve these estimates
substantially~\cite{Duerig:2016dvi}.

\section{Conclusion}
\label{sec:conclusion}

In this paper, we have described the potential of the ILC to improve
precision electroweak measurements and other precision probes of the
Standard  Model.  

For precision electroweak measurements, we have shown that the capabilities of the ILC are
very powerful, both using data from running at 250~GeV and especially
from a dedicated GigaZ stage of running at the $Z$ pole.   The
availability of polarised beams, both for electrons and for positrons,
is an important part of this story.  With few exceptions, the limiting
factor in precision electroweak measurements is not statistics but
rather the control of systematic errors.  We have explained how the
ILC experiments will  control the effective  polarisation of beams to 
the level of $(3-5)\times 10^{-4}$ and will transfer this level of
precision to the most important electroweak observables. 

We have also reviewed and updated the ILC capabilities for studies of 
fermion pair production, including $b$ and $t$ quark production.   The
ILC will determine the top quark mass to a precision of $2\times
10^{-4}$.  It will also give strong bounds on the presence of $Z'$
resonances and contact interactions due to new physics at high
energies.  The availability of polarised beams will make these
constraints very specific in terms of the flavor and helicity
structure of the operators probed---or, in the event that a deviation
from the SM is discovered, will make the origin of the corrections to
the SM very clear.

Finally, we have reviewed the ILC capabilities for the measurement of 
Higgs boson couplings, including the top quark Yukawa coupling and 
the Higgs boson self-coupling.  As we learn more about the ILC capabilities
for other measurements, the projection that the ILC will measure the
 couplings of the Higgs boson visible in Higgs decay to precisions below 1\%
remains robust. 

\newpage

\noindent {\bf Acknowledgements}

We thank the members of the Higgs@Future Colliders working group and
the BSM working group of the current study for the European Strategy
for Particle Physics for stimulating the writing of this paper, and
for many discussions of the analyses presented here.  We are
especially grateful to Juan Alcaraz, Jorge de Blas, Beate Heinemann, Aleandro
Nisati, Andrea Wulzer, and Kaoru Yokoya
for discussions of this material and to
Alain Blondel and Patrick Janot for useful correspondence. 
We are grateful to many funding agencies around the world for
supporting the research reported here and the preparation of this
document.  Among these are: the Deutsche Forschungsgemeinschaft
 (DFG, German Research 
Foundation) under Germany‘s Excellence Strategy
 – EXC 2121 "Quantum Universe" – 390833306, the Generalitat Valenciana under grant 
PROMETEO 2018/060 and the Spanish national program for particle 
physics under grant FPA2015-65652-C4-3-R, the Japan Society for the
 Promotion of Science (JSPS) under
 Grants-in-Aid for Science Research 15H02083, 16H02173,  and 16H02176,
 the US National Science Foundation under grant PHY-1607262,  and 
 the  US Department of Energy under   
 contracts DE--AC02--76SF00515, DE--AC05--06OR23177, and  DE--SC0017996.

\newpage

\appendix

\section{ILC projected uncertainties on precision electroweak\\ 
  observables and Higgs boson couplings}

Table~\ref{tab:PEWresults}:  Projected precision of precision electroweak quantities
  expected from the ILC.  Precisions are given as {\it relative}
  errors ($\delta A = \Delta A/A$) in units of $10^{-4}$.  The first
  two columns list the quantities and their SM values.   The third
  column gives the current uncertainty~\cite{Tanabashi:2018oca}.   The fourth and
  fifth columns show the statistical and systematic errors expected
  from a GigaZ run of the ILC.    The fifth and sixth columns show the 
statistical and systematic errors expected from 250~GeV running of the
ILC from  measurements of $\ee\to Z\gamma$.   In cases where the
statistical uncertainty is negligible with respect to the systematic
uncertainty, or vice versa, the smaller  element of the table is
left blank.   Boxes in which there is no improvement are marked ``-''.
 Footnote 
symbols show the source of the dominant systematic error:  $^\dagger$
acceptance; $^\circ$  energy scale;
$^*$ beam polarisation; $^+$ flavor tag; $^\#$ multiple effects,
improved from SLC.
The errors on $g_L$, $g_R$ are computed from those on $R$, $A$, and
$\Gamma_Z(had)$. 

Table~\ref{tab:ILCHiggs}:  Projected uncertainties in the Higgs
  boson couplings for the ILC250, ILC500, and ILC1000, with
  precision LHC input.  All values are  {\it relative}
  errors,
given in percent (\%).  The columns labelled ``full'' refer to a
  22-parameter fit including the possibility of invisible and exotic
  Higgs boson decays.   The columns labelled ``no BSM'' refer to a
  20-parameter fit including only decays modes present in the SM. 
The definition of a Higgs coupling
  uncertainty, 
for the purpose of this
  table,  is half the fractional uncertainty in the corresponding
  Higgs 
boson partial width. The
  bottom lines give, for reference, the projected uncertainties in the
  Higgs boson total width and the 95\% confidence limits on the Higgs
  boson invisible width.
   The analysis, which applies Effective Field Theory as described in
   the text,  is highly model-independent.  The ILC250 does 
not have direct sensitivity to the $htt$ and $hhh$ couplings; 
thus no  model-independent  values are given in these lines.  The
projections presented for these couplings in the ILC500 and ILC500
columns are based on 1-parameter fits to the $ht\bar{t}$ and $hh$
production cross sections.

\begin{table}[h!]
\begin{center}
  \begin{tabular}{cc|c|cc|cc}
    \hline
    Quantity & Value &  current &  \multicolumn{2}{c}{GigaZ}     & 
  \multicolumn{2}{c}{ILC250}    \\ 
 &   &    $\delta[10^{-4}]$&
                                                 $\delta_{stat}[10^{-4}]$&
               $\delta_{sys}[10^{-4}]$ & $\delta_{stat}[10^{-4}]$ & $\delta_{sys}[10^{-4}]$ \\ 
   \hline
   boson properties \\ \hline
   $m_W$     &     80.379 &  1.5 &    -    &    -  &   &  0.3 $^\circ$ \\
   $m_Z$      &     91.1876   & 0.23  &  - & -  &  - &  -  \\
  $\Gamma_Z$   &  2.4952   &     9.4     &    & 4. $^\circ$    &   -  &  - \\
  $\Gamma_Z (had)$  &  1.7444      &  11.5     &      & 4. $^\circ$  & -
                                                                  &   - \\ \hline
    $Z$-e couplings \\  \hline
    $1/R_e$     &   0.0482  & 24.  & 2. &  5 $^\dagger$ &  5.5   & 10 $^+$ \\
   $A_e$      &     0.1513   & 139.    &    1    &  5.  $^*$    & 9.5  &   3. $^*$ \\    
    $g^e_L$ & -0.632 &  16.   &     1.0     &   3.2      & 2.8   &  7.6  \\ 
    $g^e_R$ & 0.551&    18.     &      1.0    &   3.2    & 2.9  &  7.6 \\ 
   \hline    
     $Z$-$\ell$ couplings \\  \hline 
   $1/R_{\mu}$ & 0.0482 & 16.    &  2.  & 2. $^\dagger$ &     5.5 &  10  $^+$ \\
    $1/R_{\tau}$ & 0.0482 &   22.    & 2. &  4. $^\dagger$&    5.7 &  10  $^+$ \\
    $A_{\mu}$ & 0.1515 &   991. & 2. &  5  $^*$  &
                                          54.&  3.  $^*$ \\
   $A_{\tau}$ & 0.1515  &   271.  &   2.  &   5.  $^*$  &
                                         57. & 3 $^*$ \\
    $g^\mu_L$ & -0.632 &  66.   &     1.0      &     2.3   & 4.5 & 7.6\\ 
    $g^\mu_R$ & 0.551&     89.    &     1.0     &   2.3    &5.5 &  7.6  \\ 
    $g^\tau_L$ & -0.632 &    22.   &      1.0   &    2.8    & 4.7 & 7.6 \\ 
    $g^\tau_R$ & 0.551 &   27.     &      1.0      &    3.2   & 5.8
                                                          &  7.6 \\  \hline
   $Z$-$b$ couplings  \\ \hline
   $R_b$ & 0.2163 & 31.   &  0.4 &   7.  $^\#$  & 3.5  &   10 $^+$  \\ 
    $A_b$ & 0.935 &   214.      &  1.   &  5.  $^*$ &  5.7  &   3 $^*$\\ 
    $g^b_L$ & -0.999 & 54.&0.32 &4.2  & 2.2& 7.6\\ 
    $g^b_R$ & 0.184 &1540 &7.2 &  36.& 41.& 23.\\  \hline
    \hline
  $Z$-$c$ couplings \\ \hline  
  $R_c$ & 0.1721 & 174.  & 2.  &  30 $^\#$   &   5.8  & 50 $^+$  \\
    $A_c$ & 0.668 & 404.  & 3.   &  5$^*\oplus 5^\#$ &  21. &  3 $^*$ \\
    $g^c_L$ & 0.816 & 119. &1.2 & 15. &5.1  &  26. \\ 
    $g^c_R$ & -0.367 & 416. &3.1  &  17. &21. & 26. \\ 
    \hline
  \end{tabular}
\end{center}
\caption{Projected precision of precision electroweak quantities
  expected from the ILC.  Precisions are given as {\it relative}
  errors ($\delta A = \Delta A/A$) in units of $10^{-4}$. Please see
  the text of Appendix A  for further  explanation of this
  table.}
 \label{tab:PEWresults}
 \end{table}

\begin{table}[h!]
\begin{center}
\begin{tabular}{l|cc|cc|cc}
      &  \multicolumn{2}{c}{ ILC250 }     &  
\multicolumn{2}{c}{ ILC500}  &
               \multicolumn{2}{c}{ ILC1000 }\\ 
coupling & full & no BSM & full & no BSM & full & no BSM \\ \hline 
$hZZ$            &             0.48 & 0.38 &            0.35 &  0.20    &
                 0.34  &  0.16   \\ 
$hWW$            &          0.48 & 0.38  &            0.35  &  0.20      &                                                
                              0.34 &  0.16 \\ 
 $hbb$            &     0.99  & 0.80  & 0.58 &0.43   &  0.47 & 0.31 \\ 
$h\tau\tau$    &          1.1 & 0.95 &   0.75 & 0.64 & 0.63 & 0.52 \\ 
$hgg$                  &  1.6 &  1.6    &   0.96  & 0.92    
 & 0.67 &  0.59 \\ 
$hcc$         &   1.8  &  1.8&  1.2   &   1.1 &    0.79 &   0.72  \\ 
$h\gamma\gamma$ &  1.1  & 1.1 &   1.0  &  0.97 & 0.94  & 0.89 \\ 
$h\gamma Z$     &  8.9 &  8.9 &   6.5 &  6.5 &  6.4  & 6.4 \\
$h\mu\mu$ &  4.0  & 4.0 &  3.8  &  3.8 &     3.4  &  3.4     \\ 
$htt$  &   ---     &   --- &   6.3 & 6.3  &      1.6 & 1.6    \\ 
$hhh$  &  ---    &  ---& 27 &  27 &    10  & 10  \\ \hline 
$\Gamma_{tot}$ & 2.3 & 1.3 & 1.6 & 0.70 &  1.4  &  0.50\\  
$\Gamma_{inv}$ &   0.36 & ---   & 0.32 & --- &  0.32  & --- \\  \hline
\end{tabular}
\end{center}
\caption{Projected uncertainties in the Higgs
  boson couplings for the ILC250, ILC500, and ILC1000, with
  precision LHC input.  All values are  {\it relative}
  errors,
given in percent (\%).   The columns labelled ``full'' refer to a
  22-parameter fit including the possibility of invisible and exotic
  Higgs boson decays.   The columns labelled ``no BSM'' refer to a
  20-parameter fit including only decays modes present in the SM. 
Please see
  the text of Appendix A  for further explanation of this table.}
\label{tab:ILCHiggs}  
\end{table}

\newpage

\section{Uncertainties on observables used as inputs to the ILC
  Higgs boson coupling projections}

Table~\ref{tab:PEW}:  Values and uncertainties for precision electroweak
  observables used in this paper.  Current  values are taken from
  \cite{Tanabashi:2018oca}, except for the averaged value of $A_\ell$, which
  corresponds to the averaged value of 
  $\sin^2\theta_{eff}$ in  \cite{ALEPH:2005ab}.  The best
  fit values are those of the fit in \cite{Tanabashi:2018oca}.  For the
  purpose of fitting Higgs boson couplings as described in Sec.~\ref{sec:SMEFT},
  we use expected  improvements from the ILC  in the uncertainty on the Higgs
  mass, from~\cite{Bambade:2019fyw}, and improvements on the precision
  electroweak observables and $W$ properties explained in this paper.

Table~\ref{tab:higgserrors}   Projected statistical errors, quoted as
relative errors in \%, for Higgs boson 
measurements input to our fits. The errors are 
quoted for luminosity samples of 250~fb$^{-1}$
  for $\ee$ beams with -80\% electron polarisation and +30\% positron
  polarisation, in the top half of the table, and with  +80\% electron polarisation and -30\% positron
  polarisation, in the bottom half of the table---except that the
  values in the column for 1000~GeV are given for $\mp 80\%/\pm 20\%$
  $e^-$/$e^+$ polarization.
  Except for the first and last segments of  each set, these are measurments
  of  $\sigma \cdot BR$, relative to the Standard Model
  expectation.
The top lines gives the error for the total cross section relative to
the Standard Model and  the 95\% confidence upper limit on the branching ratio for
Higgs to invisible decays.  The bottom lines in each half give the
expected errors on the $a$ and $b$ parameters and their correlation
(all in \%)  for $\ee\to Zh$ (see \cite{Barklow:2017awn}).
   All error estimates in this table are
based on full simulation, or on the extrapolation of full simulation
results to different energies or polarization settings.

Table~\ref{tab:WWerrors}:   Projected statistical errors, in \%, for $\ee\to W^+W^-$
measurements input to our fits.   The errors are derived from a
3-parameter fit, assuming $SU(2)\times U(1)$ invariance, to the
quantities $g_{1Z}$, $\kappa_A$, $\lambda_A$ characterizing triple
gauge bosons anomalous couplings.   These errors are then interpreted
in SMEFT according to the prescription given in \cite{Barklow:2017awn}. The errors are 
quoted for luminosity samples of 500~fb$^{-1}$   divided equally
between beams with 
 -80\% electron polarisation and +30\% positron
  polarisation and  beams with  +80\% electron polarisation and -30\% positron
  polarisation. The last three lines give the correlation
  coefficients, also in \%.   All error estimates in this table are
based on full simulation, or on the extrapolation of full simulation
results to different energies.

\vskip 1.0in

\begin{table}[h!]
\begin{center}
\begin{tabular} {lccccc}
Observable  &    current value &   current $\sigma$ &  ILC250 $\sigma$
  & GigaZ $\sigma$ & 
 SM best fit value \\ \hline
$\alpha^{-1}(m_Z^2) $&        128.9220            &  0.0178       &    &
&     (same)             \\ 
$G_F$   ($10^{-10}$ GeV$^{-2}$)    &     1166378.7        & 0.6
                                                    &   &
&       (same)                           \\
$m_W$  (MeV) &          80385                   &  15   &   2.5    &  &
80361  \\
$m_Z$  (MeV)  &          91187.6                  &   2.1   & &  &
91.1880 \\ 
$m_h$   (MeV) &      125090    &   240    &   15 &   &   125110   \\ 
$A_\ell $ &         0.14696      &      0.0013    & 0.00015
  &  0.000075
&          0.147937  \\          
$\Gamma_e$  (MeV)   &         83.919                  &   0.105
&  0.072  &  0.047
&        83.995                            \\
$\Gamma_Z$ (MeV)  &         2495.2                   &   2.3   &  
&  1.   &       2.4943                      \\
$BR(W\to e\nu)$ (\%) &    10.71               &  0.16   &  0.011
     &  &   10.86   \\
\end{tabular}
\end{center}
\caption{Values and uncertainties for precision electroweak
  observables used in this paper.  Please see
  the text of Appendix B  for the explanation of this table.}
\label{tab:PEW}
\end{table}

\begin{table}[h!]
\begin{center}
\begin{tabular} {l|cc|cc|cc|c}
\multicolumn{8}{l}{-80\% $e^-$ polarisation, +30\% $e^+$  polarisation: }\\  \hline
 & 250 GeV &  & 350 GeV&
 & 500 GeV &  & 1000 GeV \\
 &  $Zh$ & $ \nu\bar\nu h$  &  
$ Zh$ & $\nu\bar\nu h$ & $Zh$ &
 $ \nu\bar\nu h$    &
 $ \nu\bar\nu h$\\ 
\hline
$\sigma$ &    2.0     &    &1.8  &  &  4.2   &     \\  \hline
$h\to invis.$ &  0.86   &  &1.4 &   &   3.4 &     \\
\hline
$h\to b\bar b$ &   1.3 &  8.1 & 1.5  &
                                                                      1.8
                                & 2.5  &  0.93 & 1.0 \\ 
$h\to c\bar c$ &  8.3 &   &11 & 19  &   18 & 8.8& 6.2 \\ 
$h\to gg$ &  7.0 &  &8.4  & 7.7 & 15  &  5.8  &4.6\\
$h\to WW$ &  4.6 &   &5.6 & 5.7 
                                &  7.7   &  3.4& 3.2 \\
$h\to \tau\tau$ & 3.2 &   &4.0 & 16 &  6.1
 &  9.8& 6.2\\
$h\to ZZ$ & 18 &   & 25 & 20 & 35  &
                                                                       12 & 8.2   \\ 
$h\to \gamma\gamma$ & 34 &   &39 &  45 &  47
 &  27& 17 \\ 
$h\to \mu\mu$ & 72 &   &87&  160\
                                                               $^*$ &
                                                                      120\
                                                                      $^*$
 &  100 & 62 \\
\hline\hline
$a$  &  7.6   &     &2.7 &   &  4.0   &  \\
$b$ &  2.7  &    & 0.69 &    & 0.70   &   \\
$\rho(a,b)$  &  -99.17 &   & -95.6
                  &  & -84.8 &   \\ \hline \hline
\multicolumn{8}{l}{+80\% $e^-$ polarisation, -30\% $e^+$  polarisation: }\\  \hline
 & 250 GeV   &  & 350 GeV & & 500 GeV  &  & 1000 GeV \\
 &  $Zh$ & $ \nu\bar\nu h$  &  
$ Zh$ & $\nu\bar\nu h$ & $Zh$ &
 $ \nu\bar\nu h$ &  $ \nu\bar\nu h$\\ 
\hline
$\sigma$   &    2.0     &    &1.8  &  &  4.2   &    & \\  \hline
$h\to invis.$ &  0.61   &  &1.3 &   &   2.4 &    & \\
\hline
$h\to b\bar b$  &   1.3 &  33 & 1.5  &  7.5  & 2.5  &  3.8 & 3.7 \\ 
$h\to c\bar c$  &  8.3 &   &11 & 79  &   18 & 36 & 23\\ 
$h\to gg$  &  7.0 &  &8.4  & 32 & 15  &  24 &   17\\
$h\to WW$  &  4.6 &   &5.6 & 24  &  7.7   &  14 & 12\\
$h\to \tau\tau$  & 3.2 &   &4.0  & 66 &  6.1  &  40 & 23\\
$h\to ZZ$  & 18 &   & 25 & 81 & 35  & 48  &  30 \\ 
$h\to \gamma\gamma$  & 34 &   &39 &  180 &  47 &  110 &62\\ 
$h\to \mu\mu$  & 72 &   &87 &  670 &  120 &  420 & 230  \\
\hline\hline
$a$ &  9.1   &     &3.1   &   &  4.2   &  \\
$b$ &  3.2  &    & 0.79    &    & 0.75   &   \\
$\rho(a,b)$  &  -99.39 &   & -96.6  
                  &  & -86.5 &   \\ 
\end{tabular}
\caption{Projected statistical errors, quoted as
relative errors in \%,  for Higgs boson 
measurements input to our fits. The errors are 
quoted for luminosity samples of 250~fb$^{-1}$
  for $\ee$ beams with -80\% electron polarisation and +30\% positron
  polarisation, in the top half of the table, and with  +80\% electron polarisation and -30\% positron
  polarisation, in the bottom half of the table.  Please see
  the text of Appendix B  for further  explanation of this table.}
  \label{tab:higgserrors}
\end{center}
\end{table}

\begin{table}[h!]
\begin{center}
\begin{tabular} {lccccccc}
 & 250 GeV   &  & 350 GeV & & 500 GeV  & &1000 GeV \\
 &  $ W^+ W^-$  &  &  $ W^+ W^-$  &  &  $ W^+ W^-$
                                  &     &   $ W^+ W^-$
 \\\hline
$g_{1Z}$ & 0 .062 &   &0.033  &  &0.025  &  & 0.0088\\
$\kappa_A $ & 0.096   &   & 0.049   &   & 0.034 &  & 0.011 \\
$\lambda_A$ &   0.077  &   &0.047    &  &0.037  &  & 0.0090\\ \hline\hline
$\rho(g_{1Z},\kappa_A)$ & 63.4   &   &63.4    &  & 63.4 &   &  63.4\\
$\rho(g_{1Z},\lambda_A)$ &  47.7 &   & 47.7   &
  &47.7  &  & 47.7 \\
$\rho(\kappa_A,\lambda_A)$ & 35.4    &   & 35.4   
 &  & 35.4  &  & 35.4
\end{tabular}
\caption{Projected statistical errors, in \%, for $\ee\to W^+W^-$
measurements input to our fits. The errors are 
quoted for luminosity samples of 500~fb$^{-1}$   divided equally
between beams with 
 -80\% electron polarisation and +30\% positron
  polarisation and  brams with  +80\% electron polarisation and -30\% positron
  polarisation.  Please see
  the text of Appendix B  for further  explanation of this table.}
\label{tab:WWerrors}
\end{center}
\end{table}

\end{document}